\documentclass{aa}  
\usepackage{graphicx}
\usepackage{txfonts}
\bibpunct{(}{)}{;}{a}{}{,} % to follow the A&A style

\usepackage{color}
\definecolor{c1}{rgb}{0.1,0.0,0.5}  

\definecolor{c2}{rgb}{0.9,0.0,0.5}

\begin{document} 

   \title{Planet-star interactions with precise transit timing}

   \subtitle{IV. Probing the regime of dynamical tides for GK host stars}

   \author{G.~Maciejewski\inst{1}
          \and
          J.~Golonka\inst{1}
          \and
          M.~Fern\'andez\inst{2}
          \and
          J.~Ohlert\inst{3,4}
          \and
          V.~Casanova\inst{2}
          \and
          D.~P\'erez~Medialdea\inst{2}
          }

   \institute{Institute of Astronomy, Faculty of Physics, Astronomy and Informatics,
              Nicolaus Copernicus University in Toru\'n, Grudziadzka 5, 87-100 Toru\'n, Poland\\
              \email{gmac@umk.pl}
         \and
             Instituto de Astrof\'isica de Andaluc\'ia (IAA-CSIC), 
             Glorieta de la Astronom\'ia 3, 18008 Granada, Spain
         \and
             Michael Adrian Observatorium, Astronomie Stiftung Trebur, 
             65428 Trebur, Germany
         \and
             University of Applied Sciences, Technische Hochschule Mittelhessen, 
             61169 Friedberg, Germany
             }
   \authorrunning{G.~Maciejewski et al.}
   \date{Received 3 September 2024 ; accepted 23 October 2024}
  
  \abstract
   {Giant exoplanets on 1--3 day orbits, known as ultra-hot Jupiters, induce detectable tides in their host stars. The energy of those tides dissipates at a rate related to the properties of the stellar interior. At the same time, a planet loses its orbital angular momentum and spirals into the host star. The decrease of the orbital period is empirically accessible with precise transit timing and can be used to probe planet-star tidal interactions.}
   {Statistical studies show that stars of GK spectral types, with masses below 1.1 $M_{\odot}$, are depleted in hot Jupiters. This finding is evidence of tidal orbital decay during the main-sequence lifetime. Theoretical considerations show that in some configurations, the tidal energy dissipation can be boosted by non-linear effects in dynamical tides, which are wave-like responses to tidal forcing. To probe the regime of these dynamical tides in GK stars, we searched for orbital period shortening for 6 selected hot Jupiters in systems with 0.8--1 $M_{\odot}$ host stars: HATS-18, HIP 65A, TrES-3, WASP-19, WASP-43, and WASP-173A.}
   {For the hot Jupiters of our sample, we analysed transit timing data sets based on mid-transit points homogeneously determined from observations performed with the Transiting Exoplanet Survey Satellite and high-quality data available in the literature. For the TrES-3 system, we also used new transit light curves we acquired with ground-based telescopes. The mid-transit times were searched for shortening of orbital periods through statistical testing of quadratic transit ephemerides. Theoretical predictions on the dissipation rate for dynamical tides were calculated under the regimes of internal gravity waves (IGWs) undergoing wave breaking (WB) in stellar centres and weak non-linear (WNL) wave-wave interactions in radiative layers. Stellar parameters of the host stars, such as mass and age, which were used in those computations, were homogeneously redetermined using evolutionary models with the Bayesian inference.}
   {We found that transit times follow the refined linear ephemerides for all ultra-hot Jupiters of our sample. Non-detection of orbital decay allowed us to place lower constraints on the tidal dissipation rates in those planet-star systems. In three systems, HATS-18, WASP-19, and WASP-43, we reject a scenario with total dissipation of IGWs. We conclude that their giant planets are not massive enough to induce wave breaking. Our observational constraints for HIP 65A, TrES-3, and WASP-173A are too weak to probe the WB regime. Calculations show that wave breaking is not expected in the former two, leaving the WASP-173A system as a promising target for further transit timing observations. The WNL dissipation was tested in the WASP-19 and WASP-43 systems, showing that the theoretical dissipation rates are overestimated by at least one order of magnitude. For the remaining systems, decades or even centuries of transit timing measurements are needed to probe the WNL regime entirely. Among them, TrES-3 and WASP-173A have the predicted WNL dissipation rates that coincide with the values obtained from gyrochronology.}
   {Tidal dissipation in the GK stars of our sample is not boosted by wave breaking in their radiative cores, preventing their giant planets from rapid orbital decay. Weakly non-linear tidal dissipation could drive orbital shrinkage and stellar spin-up on Gyr timescales. Although our first results suggest that theory might overestimate the dissipation rate and some fine-tuning would be needed for at least a fraction of planet-star configurations, some predictions coincide intriguingly with the gyrochronological estimates. We identify the WASP-173A system as a promising candidate for exploring this problem in the shortest possible time of the coming decades.}


   \maketitle
%
%-------------------------------------------------------------------

\section{Introduction}\label{Sect:Intro}

Massive planets on extremely tight orbits, so-called ultra-hot Jupiters, are considered unique tools for probing some of the properties of stellar interiors via tidal interactions. Population studies show a deficit of such planets around Sun-like stars advanced in their evolution on the main sequence \citep{2023AJ....166..209M}. This finding is in line with an observation that the population of hot Jupiters' host stars is relatively young \citep{2019AJ....158..190H,2023PNAS..12004179C}, implying that these planets effectively lose their orbital angular momentum on a timescale of a stellar lifetime at the main sequence stage. The hot-Jupiter hosts rotate faster than stars without those planetary companions, showing that they were tidally spun-up \citep{2021ApJ...919..138T}. According to theoretical expectations, this process is governed by the dissipation of tidal energy in stellar interiors. Statistical studies provide a rough estimation of the scale of this effect \citep[e.g.][]{2012ApJ...751...96P}, but its magnitude in individual star-planet systems might vary significantly, depending on the specific physical properties of particular systems \citep{2018AJ....155..165P,2020MNRAS.498.2270B}. The efficiency of tidal dissipation is characterised by the modified tidal quality factor $Q'_{\star}$, which is proportional to the ratio of the energy stored in a tide and the energy dissipated in a single tidal cycle \citep[e.g.][]{2020MNRAS.498.2270B}. This quantity is directly related to the change of the planet's orbital period $\dot{P}_{\rm orb}$, an observable accessible with the method of precise transit timing.

There are two types of manifestations of tides affecting the stellar interior. One of them is equilibrium tides, quasi-hydrostatic ellipsoidal deformations induced by the gravity of a planetary companion. They dissipate the tidal energy thanks to the effective viscosity of a tidal flow in a convection layer \citep{1966AnAp...29..489Z}. They were found to be negligibly weak in typical planet-star configurations on the main sequence \citep{2020MNRAS.498.2270B} and are, therefore, skipped in this study. Dynamical tides are another phenomenon induced by tides. They are low-frequency waves resonantly excited by tidal forcing and operate in stellar layers supporting oscillations. They manifest as inertial waves in convective zones and internal gravity waves (IGWs) in radiative layers. The former might be an efficient driver for tidal dissipation in configurations where the tidal period is longer than half the rotation period of the host star. In practice, such waves can be significant in specific planet-star systems where host stars are close to spin-orbit synchronisation. As that does not apply to our sample systems, we skip them in this study. The internal gravity waves, in turn, are excited at a radiative-convective boundary and propagate towards the stellar centre. In stars with non-convective cores, their amplitudes may increase due to geometric focusing, leading to strongly non-linear effects and full damping by wave breaking \citep[WB,][]{2010MNRAS.404.1849B}. The high efficiency of tidal dissipation in this scenario is expected to produce rapid orbital decay, which is easily accessible with precise transit timing measurements over the course of several decades. These waves can also form a sea of secondary modes through wave-wave interactions in the regime of weak non-linearity \citep[WNL,][]{2016ApJ...816...18E,2024ApJ...960...50W}. Their energy could be efficiently dissipated by radiative damping \citep{1975A&A....41..329Z}, causing moderate rates of orbital decay. Interestingly, most of the ultra-hot Jupiters might be in this regime; for those in the most favourable configurations, the observational effects could also be detectable over a few decades \citep{2024ApJ...960...50W}.

As transit timing data sets build up with time, these theoretical predictions gradually become available for observational verification, triggering a series of attempts to detect orbital decay of hot Jupiters \citep[e.g.][]{2022AJ....164..220H,2024ESS.....560501W,2024PSJ.....5..163A}. Our previous study \citep{2022A&A...667A.127M} placed observational constraints on $Q'_{\star}$ values, ruling out wave breaking in 3 systems with host stars being more massive than 1.1 $M_{\odot}$. This negative result was unsurprising because ultra-hot Jupiters in those systems have masses below the critical value $M_{\rm crit}$ required for wave breaking, and the host stars likely have convective cores \citep{2020MNRAS.498.2270B}. This paper directs our interest to a sample of six GK host stars with masses ranging from 0.8 to 1.0 $M_{\odot}$ and being orbited by ultra-hot Jupiters: HATS-18 \citep{2016AJ....152..127P}, HIP 65A \citep{2020AA...639A..76N}, TrES-3 \citep{2007ApJ...663L..37O}, WASP-19 \citep{2010ApJ...708..224H}, WASP-43 \citep{2011AA...535L...7H}, and WASP-173 A \citep{2019MNRAS.482.1379H}. They are expected to have radiative cores during their evolution stage at the main sequence, potentially allowing for the geometrical focusing of gravity waves in their centres. For three of them -- HATS-18, WASP-19, and WASP-173A, their planetary companions were predicted to have masses high enough that WB could operate \citep{2020MNRAS.498.2270B}. We also address the tidal dissipation under the WNL regime, which could be another efficient driver for orbital decay in those six systems, as pointed out by \citet{2024ApJ...960...50W}. Our analysis is based on the precise timing of transits extracted from photometric time series acquired with the Transiting Exoplanet Survey Satellite \citep[TESS,][]{2015JATIS...1a4003R}, and in the case of TrES-3, enhanced with high-quality ground-based follow-up observations.

The rest of this paper is organised as follows. In Sect.~\ref{Sect:Obs}, we present the original observational material used in our study and provide details on its reduction. In Sect.~\ref{Sect:Results}, we describe the analysis process and its outcomes. In Sect.~\ref{Sect:Discussion}, we discuss our findings for the individual systems. Finally, we present conclusions in Sect.~\ref{Sect:Conclusions}.

\section{Observations and their reduction}\label{Sect:Obs}

\subsection{TESS photometric time series}\label{SSect.TESSobs}

\begin{figure}
    \includegraphics[width=\columnwidth]{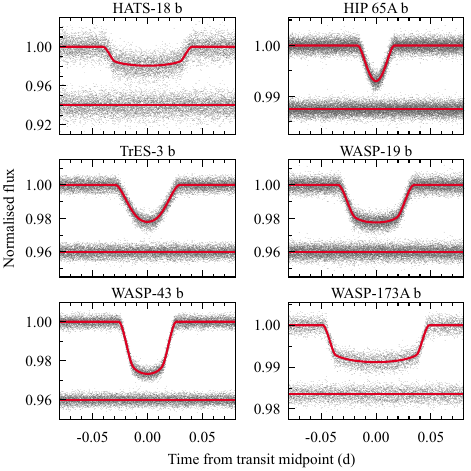}
    \caption{Phase folded transit light curves for the hot Jupiters in our sample, extracted from TESS photometry. The best-fitting models (Sect.~\ref{SSect:Results.Parameters}) are drawn with red lines, and the photometric residuals, shifted arbitrarily for clarity, are plotted below.}
    \label{fig:tesslcs}
\end{figure}

The sample systems were observed with TESS in at least 3 sectors at the time coverage ranging from 2 to 5 years. We considered only short-cadence data with an exposure time of 2 minutes. A summary of these observations is provided in Table~\ref{tab.ObsTESS}. The light curves were extracted from science frames using standard procedures implemented in the \texttt{Lightkurve} package \citep[ver.\ 2.0.11,][]{2018ascl.soft12013L}. The target pixel files, centred at the target position and $11 \times 11$ pixels ($3 \farcm 85 \times 3 \farcm 85$) wide, were downloaded from the exo.MAST portal\footnote{https://exo.mast.stsci.edu}. The aperture photometry method was applied with a mask, the shape of which was iteratively optimised to account for inter-sector changes of the stellar profile and orientation of the field of view. Instrumental trends and possible stellar variability were cleaned out by applying the Savitzky-Golay filter \citep{1964AnaCh..36.1627S} with a window width of 8 hours and in-transit and in-occultation chunks masked out. Detrended and normalised light curves were carefully inspected for sporadic outlying measurements, remaining time-correlated flux ramps, and scattered light. The corrupted data points were removed. The phase folded light curves for transits of the hot Jupiters in the individual systems are plotted in Fig.~\ref{fig:tesslcs}.

\subsection{Ground-based observations for TrES-3}\label{SSect:GObs}

\begin{figure}
    \includegraphics[width=\columnwidth]{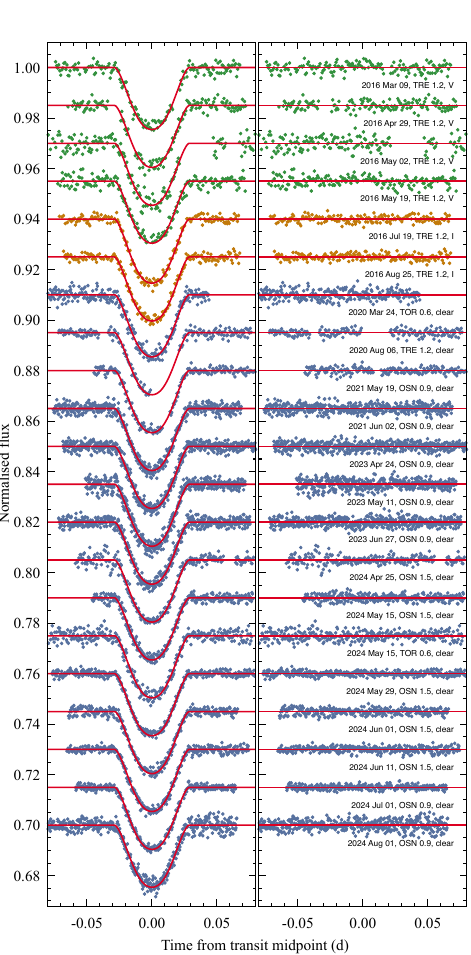}
    \caption{New ground-based transit light curves for TrES-3~b. Left: Individual photometric time series sorted by the observation date. The $V$-band data are marked with green points, the $I$-band data are marked with orange points, and the white-light (clear filter) data are marked with blue points. The best-fitting models (Sect.~\ref{SSect:Results.Parameters}) are drawn with red lines. Right: Photometric residuals from the transit model.}
    \label{fig:tr3lcs}
\end{figure}

We obtained supplementary transit photometry for TrES-3~b with ground-based instruments. Fifteen light curves were acquired between March 2020 and early August 2024 with the 1.5 m and 0.9 m Ritchey-Chr\'etien telescopes (OSN 1.5 and OSN 0.9) at the Sierra Nevada Observatory (OSN, Spain) equipped with Roper Scientific VersArray 2048B CCD cameras with pixel scales of 0.232$\arcsec$/pixel and 0.387$\arcsec$/pixel, respectively\footnote{See https://www.osn.iaa.es/en/page/ccdt150-and-ccdt90-cameras for more details.}; the 1.2 m Trebur telescope (TRE 1.2) at the Michael Adrian Observatory in Trebur (Germany) with an SBIG STL-6303 CCD camera with a pixel scale of 0.389$\arcsec$/pixel; and the 0.6 m Cassegrain photometric telescope (TOR 0.6) at the Institute of Astronomy of the Nicolaus Copernicus University in Toru\'n (Poland) with an FLI 16803 CCD camera in 2020 and a Moravian Instruments C4-16000EC CMOS camera in 2024 with pixel scales of 0.517$\arcsec$/pixel and 0.541$\arcsec$/pixel, respectively. The instruments were automatically (OSN 1.5 and OSN 0.9) or manually (TRE 1.2 and TOR 0.6) guided during each observing run to minimise field drifts. Scientific exposures were acquired in white light (a clear filter) with the optical setups being defocused to reduce flat-fielding errors and avoid saturation at longer exposure times \citep[e.g.][]{2009MNRAS.396.1023S}. The details of the individual observing runs are given in Table~\ref{tab.Obs}.

The standard data calibration and reduction procedure was applied to CCD frames using the aperture photometry method described in \citet{2022A&A...667A.127M}. In brief, the light curves were generated with \texttt{AstroImageJ} \citep{2017AJ....153...77C} after iterative optimisation of the aperture size and the ensemble of comparison stars. Detrending against time, airmass, xy matrix position, and seeing was applied, and normalisation to a flux level outside transits was done. The timestamps in the final light curves were converted to barycentric Julian dates and barycentric dynamical time $\rm{BJD_{TDB}}$ using a time conversion calculator implemented in the code.

We also carefully re-reduced the original observations from \citet{2017A&A...608A..26M}, which were acquired with TRE 1.2 in the Bessel $V$ and Cousins $I$ bands between March and August 2016. Those observations were used for atmospheric studies of TrES-3~b and not used for transit timing. We followed the abovementioned procedure, producing 6 additional transit light curves from the original science frames.

A total of 21 new ground-based light curves were used in our timing analysis for TrES-3~b. They are presented in Fig.~\ref{fig:tr3lcs} together with models obtained in Sect.~\ref{SSect:Results.Parameters}.

\section{Data analysis and results}\label{Sect:Results}

\subsection{Transit light curve modelling}\label{SSect:Results.Parameters}

The transit light curves were modelled with the \texttt{Transit Analysis Package} \citep[\texttt{TAP},][]{2012AdAst2012E..30G}. The chunks of the light curves centred at transits were extracted from the TESS data. Their length was 5 transit durations to provide out-of-transit data long enough to account for possible remnant trends. Only complete transits were considered to ensure the reliability of our results. The numbers of transits used for each system and sector are given in Table~\ref{tab.ObsTESS}. 

The \texttt{TAP} code uses an analytic transit model with the following free parameters: the orbital inclination $i_{\rm{b}}$, the semi-major axis scaled in stellar radii $a_{\rm{b}}/R_{\star}$, the ratio of planet to star radii $R_{\rm{b}}/R_{\star}$, the limb darkening coefficients (LDCs) of a quadratic law $u_{\rm 1}$ and $u_{\rm 2}$ in a given passband, times of midpoints $T_{\rm{mid}}$, and possible remnant variations in flux approximated with a second-order polynomial. In the case of grazing or almost grazing transits for HIP 65A~b and TrES-3~b, the distribution of the limb-to-centre brightness is not probed, and the photometric data do not yield reliable values of LDCs \citep{2013AA...560A.112M}. In those two systems, we interpolated LDCs from the theoretical tables of \citet{2011AA...529A..75C} and varied under the Gaussian penalty of the width of 0.1 for $u_{\rm 1}$ and 0.2 for $u_{\rm 2}$ to account to some extent for possible offsets \citep{2022AJ....163..228P}. The time-correlated noise was assessed using the wavelet-likelihood method \citep{2009ApJ...704...51C}. Ten Markov chain Monte Carlo (MCMC) walkers, each $10^6$ steps long with a 10\% burn-in phase, were employed to explore the space of the parameters. Their posterior probability distributions were used to identify the best-fitting values and their uncertainties at the 50, 15.9, and 84.1 percentiles. The results are presented in Table~\ref{tab.results} and compared with the values found in the literature. The best-fitting transit models are overplotted on the phase-folded TESS light curves shown in Fig.~\ref{fig:tesslcs}. The individual mid-transit times are collected in Table~\ref{tab.TTimes}. 

For TrES-3, the modelling procedure was also applied to our new ground-based transit light curves. The value of $R_{\rm{b}}/R_{\star}$ was found to be greater by $6\%$ at a significance of 3.3 $\sigma$, showing that our TESS data suffer from contaminating flux from nearby sources. Indeed, the crowding parameter (\texttt{CROWDSAP}), calculated by the TESS data reduction pipeline, shows that $5\%$ of the flux in the aperture is due to other sources. The remaining parameters agreed within 2.4 and 3.1 $\sigma$. For further analysis, we used the individual mid-transit times only. They are given in Table~\ref{tab.TTimes}. The transit on 15 May 2024 was observed simultaneously with two instruments. Both light curves provided independent values of the mid-point for that epoch, allowing us to check the accuracy of the absolute timing. They were found to differ by 16 seconds and agree well within their 1-$\sigma$ uncertainties.

For the sake of the homogeneity of the transit timing data sets, we redetermined the midpoints for the literature data using TAP. We followed the procedure applied to our original observational material. We considered only complete and high-quality transit light curves, which were publicly available or provided upon request. We carefully inspected them and rejected those we flagged as likely affected by red noise signatures of any origin. Below, we briefly summarise the data included in our analysis. For HATS-18~b, we used a light curve from \citet{2016AJ....152..127P}, obtained with a 1-m telescope on 22 January 2016. Its timestamps were corrected following \citet{2022MNRAS.515.3212S}, who found a discrepancy between the declared and actual time systems. We also qualified 15 light curves from \citet{2022MNRAS.515.3212S} gathered with a 1.5 m telescope between May 2017 and June 2021. For TrES-3~b, we used 36 light curves of the highest quality from \citet{2009ApJ...691.1145S}, \citet{2010MNRAS.408.1494C}, \citet{2013ApJ...764....8K}, \citet{2013IBVS.6082....1M}, \citet{2013MNRAS.428..678T}, \citet{2013MNRAS.432..944V}, \citet{2017A&A...608A..26M}, \citet{2017NewA...55...39P}, and \citet{2019AA...628A.115V}. They were acquired between March 2007 and August 2018, providing extended coverage preceding our ground-based observations and the TESS runs. We redetermined the mid-points for 42 transits of WASP-19~b ranging from December 2008 to April 2019. Those photometric time series were taken from \citet{2010ApJ...708..224H}, \citet{2011AJ....142..115D}, \citet{2011ApJ...730L..31H}, \citet{2013A&A...552A...2L}, \citet{2013MNRAS.436....2M}, \citet{2013MNRAS.428.3671T}, \citet{2020AA...636A..98C}, and \citet{2020MNRAS.491.1243P}. For WASP-43~b, we found 39 transit light curves in \citet{2012AA...542A...4G}, \citet{2013IBVS.6082....1M}, \citet{2014A&A...563A..40C}, \citet{2014A&A...563A..41M}, \citet{2015PASP..127..143R}, \citet{2016AJ....151...17J}, and \citet{2021ApJS..255...15W}, falling between December 2010 and March 2017. For WASP-173A~b, we qualified a single light curve from \citet{2019MNRAS.482.1379H}, obtained with a 1.2-m telescope on 20 October 2016. All those redetermined mid-transit times are also included in Table~\ref{tab.TTimes}. In most cases, they are consistent with the original values reported in the literature at the 1 $\sigma$ level. The median differences for the individual systems varied from 4 to 10 seconds. Our timing uncertainties were, in turn, greater by about 40\% on average.

\subsection{Transit timing}\label{SSect:Results.Timing}

\begin{figure*}
        \includegraphics[width=2\columnwidth]{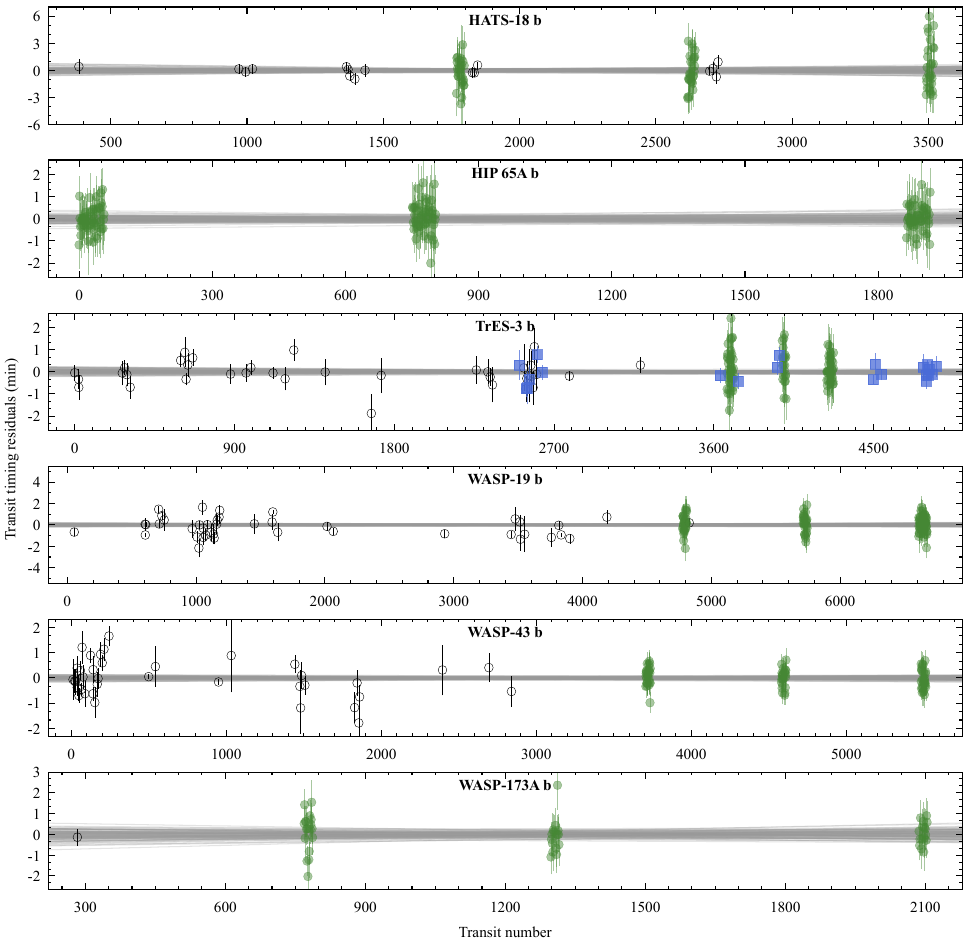}
    \caption{Transit timing residuals against the refined linear ephemerides for the planets of our sample. The filled symbols are the new determinations reported in this paper: the green and blue points come from TESS and ground-based photometry, respectively. The open circles mark the literature mid-points redetermined in this study. The uncertainties of the refined ephemerides are illustrated with samples of 100 models drawn from the Markov chains.}
    \label{fig:tt}
\end{figure*}

We adopted the methodology of transit timing analysis from \citet{2018AcA....68..371M}. We refined the linear transit ephemerides using a constant-period model as a null hypothesis in the form
\begin{equation}
  T_{\rm mid }(E) = T_0 + P_{\rm orb} \times E \, , \;
\end{equation}
where $E$ is the transit number counted from the reference epoch $T_0$, taken from the discovery paper. Then, we tried an alternative hypothesis with a quadratic ephemeris, which probes for a constantly changing orbital period. In that case, the calculated mid-transit times are expected to follow the formula 
\begin{equation}
 T_{\rm{mid}}(E)= T_0 + P_{\rm{orb}} \times E + \frac{1}{2} \frac{{\rm d} P_{\rm{orb}}}{{\rm d} E} \times E^2 \, , \;
\end{equation}
where ${{\rm d} P_{\rm{orb}}}/{{\rm d} E}$ is the change of the orbital period between succeeding transits. In both steps, the best-fitting parameters and their 1 $\sigma$ uncertainties were taken from the posterior probability distributions produced with the MCMC algorithm running 100 chains, $10^4$ steps long each, with a 10\% burn-in phase. Our model selection was based on the Bayesian information criterion (BIC), calculated for a given model as 
\begin{equation}
  {\rm BIC} = {\chi}^2 + k \ln N \, , \;
\end{equation}
where ${\chi}^2$ is the value of the chi-squared statistic, $k$ is the number of fit parameters, and $N$ is the number of data points.

%------------------ TABLE Ephemerides
\begin{table*}[h]
\caption{Parameters of the refined transit ephemerides.} 
\label{tab.ephemerides}      
\centering                  
\begin{tabular}{l c c c c c c}      
\hline\hline                
Planet & $T_0$ ($\rm{BJD_{TDB}}$)  & $P_{\rm orb}$ (d) & $\frac{{\rm d} P_{\rm{orb}}}{{\rm d} E}$ $(\times 10^{-10})$ & $N_{\rm{dof}}$ & $\chi^2$ & BIC \\
\hline
\multicolumn{7}{c}{linear ephemerides} \\
HATS-18 b   & 2457089.90564(22)  & 0.83784378(10)  & $...$ &  92 &  81.9 &  91.0 \\
HIP 65A b   & 2458326.10424(9)   & 0.98097223(8)   & $...$ & 144 &  47.4 &  57.4 \\
TrES-3 b    & 2454185.911132(68) & 1.306186250(19) & $...$ & 149 & 108.7 & 118.7 \\
WASP-19 b   & 2454775.338173(59) & 0.788838963(16) & $...$ & 159 & 270.8 & 281.0 \\
WASP-43 b   & 2455528.868585(42) & 0.813474079(12) & $...$ & 120 & 145.8 & 155.4 \\
WASP-173A b & 2457288.85945(18)  & 1.38665310(13)  & $...$ &  48 &  37.5 &  45.4 \\
\multicolumn{7}{c}{trial quadratic ephemerides} \\
HATS-18 b   & 2457089.90590(51)  & 0.83784349(52)  & $1.4 \pm 2.4$   &  91 &  81.6 &  95.3 \\
HIP 65A b   & 2458326.10423(10)  & 0.98097224(31)  & $-0.1 \pm 3.1$  & 143 &  47.4 &  62.4 \\
TrES-3 b    & 2454185.911101(94) & 1.30618630(9)  & $-0.15 \pm 0.32$ & 148 & 108.4 & 123.5 \\
WASP-19 b   & 2454775.33824(9)   & 0.78883890(7)   & $0.19 \pm 0.20$ & 158 & 269.9 & 285.1 \\
WASP-43 b   & 2455528.868588(51) & 0.81347407(6)   & $0.02\pm0.20$   & 119 & 145.8 & 160.2 \\
WASP-173A b & 2457288.85952(42)  & 1.38665296(72)  & $1.1\pm5.2$     &  47 &  37.4 &  49.2 \\
\hline                                   
\end{tabular}
\tablefoot{Uncertainties of $T_0$ and $P_{\rm orb}$ are given in the concise notation. $N_{\rm{dof}}$ is the number of degrees of freedom equal to the difference between the number of data points and the number of fit parameters.}
\end{table*}
%------------------ END OF TABLE

The new ephemerides are collected in Table~\ref{tab.ephemerides}, and the residuals against the linear models are plotted in Fig.~\ref{fig:tt}. For all planets, the quadratic ephemeris models yield higher values of BIC, so the null hypotheses cannot be rejected. In all these cases, the values of ${{\rm d} P_{\rm{orb}}}/{{\rm d} E}$ remain consistent with zero well within the 1 $\sigma$ range. These non-detections of the change of $P_{\rm orb}$ allowed us to place the lower constraints on $Q'_{\star}$ in all six systems. We found the value of ${{\rm d} P_{\rm{orb}}}/{{\rm d} E}$ at the fifth percentile of the posterior probability distribution that can be translated into the limiting value of $Q'_{\star}$ at the 95\% confidence level following  Eq.~(5) from \citet{2018AcA....68..371M}. To compute the planet-to-star mass ratio ${M_{\rm b}}/{M_{\star}}$, we used the relation
\begin{equation}
 \frac{M_{\rm b}}{M_{\star}} = 4.694 \times 10^{6} \times K_{\rm{b}} P_{\rm{orb}}^{1/3} M_{\star}^{-1/3} (\sin{i_{\rm{b}}})^{-1} \, , \;
\end{equation}
using the redetermined values of $M_{\star}$ (Sect.~\ref{SSect:Results.Qius}) and the literature values of the amplitudes of the reflex radial velocity $K_{\rm{b}}$ induced by the planets (collected in Table~\ref{tab.Stars}). The results are presented in Table~\ref{tab.ques} with other empirical constraints found in the literature and the theoretical predictions obtained in Sect.~\ref{SSect:Results.Qius}.

\begin{table}[t!]
\caption{Observational constraints (with 95\% confidence) and theoretical predictions on the values of $Q'_{\star}$.} 
\label{tab.ques}      
\centering                  
\begin{tabular}{l l}      
\hline\hline                
$Q'_{\star}$  & Type, source \\
\hline
\multicolumn{2}{c}{HATS-18 b} \\
$>$$3.5^{+0.5}_{-0.7} \times 10^{5}$  & observational, this study\\
$>$$1.29^{+0.12}_{-0.11} \times 10^{5}$  & observ., \citet{2022MNRAS.515.3212S} \\
$<$$2.6 \times 10^{5}$ & theoretical WB$^{\rm {a}}$, this study \\
$\sim$$1.1 \times 10^{5}$ & theoretical WB$^{\rm {a}}$, \citet{2020MNRAS.498.2270B} \\
$\sim$$1.2 \times 10^{5}$ & theor. WB$^{\rm {a}}$, \citet{2022MNRAS.515.3212S}\\
$\sim$$5.5 \times 10^{5}$ & theoretical WNL$^{\rm {b}}$, this study \\
$\sim$$(1-2.4) \times 10^{7}$ & spin-up$^{\rm {c}}$, \citet{2018AJ....155..165P} \\
\hline
\multicolumn{2}{c}{HIP 65A b} \\
$>$$7.6^{+0.8}_{-0.7} \times 10^{4}$ & observational, this paper \\
$(1.2 - 1.5) \times 10^{5}$ & theoretical WB$^{\rm {a}}$, this study \\
$\sim$$1.3 \times 10^{6}$ & theoretical WNL$^{\rm {b}}$, this study \\
$>$$10^{8}$ & spin-up$^{\rm {c}}$, \citet{2020AA...639A..76N} \\
\hline
\multicolumn{2}{c}{TrES-3 b} \\
$>$$2.50^{+0.12}_{-0.14} \times 10^{5}$ & observational, this paper \\
$\sim$$1.1 \times 10^{5}$ & observ., \citet{2020AJ....160...47M}\\
$>$$2.8 \times 10^{6}$ & observ., \citet{2022AJ....164..198M}\\
$(4.2-6.4) \times 10^{5}$ & theoretical WB$^{\rm {a}}$, this study \\
$\sim$$6.5 \times 10^{5}$ & theoretical WB$^{\rm {a}}$, \citet{2020MNRAS.498.2270B} \\
$\sim$$2.5 \times 10^{6}$ & theoretical WNL$^{\rm {b}}$, this study \\
$>$$4 \times 10^{6}$ & spin-up$^{\rm {c}}$, \citet{2018AJ....155..165P} \\
\hline
\multicolumn{2}{c}{WASP-19 b} \\
$>$$4.79^{+0.31}_{-0.29} \times 10^{6}$ & observational, this paper \\
$=$$(5.0 \pm 1.5) \times 10^{5}$ & observational, \citet{2020AJ....159..150P} \\
$=$$(7 \pm 1) \times 10^{5}$ & observational, KP23\\
$>$$(1.23 \pm 0.23) \times 10^{6}$ & observational, \citet{2020MNRAS.491.1243P} \\
$>$$(1.26 \pm 0.10) \times 10^{6}$ & observational, \citet{2022AA...668A.114R} \\
$<$$0.9 \times 10^{5}$ & theoretical WB$^{\rm {a}}$, this study \\
$\sim$$(0.6-0.8) \times 10^{5}$ & theoretical WB$^{\rm {a}}$, \citet{2020MNRAS.498.2270B} \\
$\sim$$4.2 \times 10^{5}$ & theoretical WNL$^{\rm {b}}$, this study \\
$\sim$$(6.5-8.1) \times 10^{6}$ & spin-up$^{\rm {c}}$, \citet{2018AJ....155..165P} \\
\hline
\multicolumn{2}{c}{WASP-43 b} \\
$>$$1.15^{+0.53}_{-0.50} \times 10^{6}$ & observational, this paper \\
$\sim$$10^{5}$ & observational, \citet{2016AJ....151...17J} \\
$>$$10^{5}$ & observational, \citet{2016AJ....151..137H} \\
$>$$1.5 \times 10^{5}$ & observational, \citet{2018ChAA..42..101S} \\
$>$$(2.1 \pm 1.4) \times 10^{5}$ & observational, \citet{2020AJ....159..150P} \\
$>$$(4.0 \pm 1.2) \times 10^{5}$ & observational, \citet{2021AJ....162..210D} \\
$(0.8-1.0) \times 10^{5}$ & theoretical WB$^{\rm {a}}$, this study \\
$\sim$$1.3 \times 10^{5}$ & theoretical WB$^{\rm {a}}$, \citet{2020MNRAS.498.2270B} \\
$\sim$$5.9 \times 10^{5}$ & theoretical WNL$^{\rm {b}}$, this study \\
$\sim$$(2.6-6.1) \times 10^{7}$ & spin-up$^{\rm {c}}$, \citet{2018AJ....155..165P} \\
\hline
\multicolumn{2}{c}{WASP-173A b} \\
$>$$1.50^{+0.19}_{-0.16} \times 10^{5}$  & observational, this paper \\
$<$$1.5 \times 10^{6}$ & theoretical WB$^{\rm {a}}$, this study \\
$\sim$$2.9 \times 10^{6}$ & theoretical WNL$^{\rm {b}}$, this study \\
$\sim$$1.6 \times 10^{6}$ & spin-up$^{\rm {c}}$, LB19 \\
\hline                                   
\end{tabular}
\tablefoot{$^{\rm {a}}$ For dissipation of internal gravity waves in radiation zones if wave breaking occurs. $^{\rm {b}}$ For dissipation of weakly nonlinear dynamical tides, interpolated from Table~4 of \citet{2024ApJ...960...50W}. $^{\rm {c}}$ From tidal spin-up of the host star. References: KP23 -- \citet{2023Univ...10...12K}, LB19 -- \citet{2019ApJS..240...13L}}
\end{table}

\subsection{Predictions for $Q'_{\star}$ in the regime of dynamical tides}\label{SSect:Results.Qius}

Since the dissipation efficiency for dynamical tides depends on systemic and stellar parameters, we homogeneously redetermined the hosts' effective temperatures $T_{\rm eff}$, surface gravity $\log g$, metalicity [Fe/H], masses $M_{\star}$, radii $R_{\star}$, luminosities $L_{\star}$, ages, and interstellar reddenings in the $V$ band $A_{\rm V}$. We employed the package {\tt isochrones} \citep{2015ascl.soft03010M}, which uses a grid of the MESA \citep[Modules for Experiments in Stellar Astrophysics,][]{2011ApJS..192....3P, 2013ApJS..208....4P, 2015ApJS..220...15P, 2018ApJS..234...34P, 2019ApJS..243...10P, 2023ApJS..265...15J} Isochrones and Stellar Tracks models \citep[MIST,][]{2016ApJS..222....8D,2016ApJ...823..102C} to interpolate the physical parameters from observed data. The implemented Bayesian inference tool {\tt MultiNest} \citep{2008MNRAS.384..449F,2009MNRAS.398.1601F,2019OJAp....2E..10F} was used to explore the space of model parameters and analyse the likelihood. As the inputs, we took the stellar atmospheric parameters inferred from pure spectral analysis, which are available in the literature: $T_{\rm eff}$, $\log g$ (if available), and [Fe/H]. We collected their values in Table~\ref{tab.Stars}. We also used the parallax and the blue $G_{\rm BP}$, green $G$, and red $G_{\rm RP}$ apparent magnitudes from the Gaia Data Release 3 catalogue \citep{2023A&A...674A...1G}. To avoid inconsistencies caused by combining heterogeneous data, we refrained from extending multi-wavelength information by 2MASS $JHK_{\rm s}$ \citep{2003yCat.2246....0C} and WISE $W1-W3$ \citep{2012yCat.2311....0C} fluxes. Since our transit models do not account for flux contamination, we refrained from using a mean stellar density, which can be calculated from $a_{\rm{b}}/R_{\star}$ and $P_{\rm{orb}}$ derived from the photometric transits. We also did not place any constraints on the interstellar extinction. The best-fitting parameters and their uncertainties were derived from the posterior probability distributions at the 50, 15.9, and 84.1 percentiles. They are presented in Table~\ref{tab.StellarPars}. 

%------------------ TABLE STELLAR PARs
\begin{table*}[h]
\caption{Redetermined stellar parameters for the host stars.} 
\label{tab.StellarPars}      
\centering                  
\begin{tabular}{l c c c c c c c c}      
\hline\hline                
Star      & $T_{\rm eff}$ (K) & $\log g_{\star}$ (cgs) & [Fe/H] & $M_{\star}$ $(M_{\odot})$ & $R_{\star}$ $(R_{\odot})$ & $L_{\star}$ $(L_{\odot})$ & Age (Gyr) &  $A_{\rm V}$ (mag)\\
\hline
HATS-18   & $5610^{+88}_{-95}$ & $4.447^{+0.029}_{-0.033}$ & $0.23^{+0.07}_{-0.07}$ & $1.024^{+0.043}_{-0.052}$ & $1.001^{+0.016}_{-0.014}$ & $0.892^{+0.053}_{-0.049}$ & $4.0^{+3.4}_{-2.5}$ & $0.14^{+0.08}_{-0.08}$ \\
HIP 65A   & $4647^{+40}_{-36}$ & $4.566^{+0.013}_{-0.007}$ & $0.250^{+0.052}_{-0.045}$ & $0.776^{+0.016}_{-0.009}$ & $0.759^{+0.004}_{-0.004}$ & $0.242^{+0.009}_{-0.008}$ & $10.8^{+1.7}_{-3.2}$ & $0.291^{+0.066}_{-0.062}$ \\
TrES-3    & $5605^{+60}_{-56}$ & $4.555^{+0.014}_{-0.017}$ & $-0.20^{+0.06}_{-0.06}$ & $0.882^{+0.021}_{-0.026}$ & $0.821^{+0.005}_{-0.005}$ & $0.600^{+0.023}_{-0.021}$ & $2.6^{+2.1}_{-1.7}$ & $0.147^{+0.050}_{-0.045}$ \\
WASP-19   & $5544^{+81}_{-70}$ & $4.384^{+0.025}_{-0.020}$ & $0.041^{+0.076}_{-0.072}$ & $0.915^{+0.038}_{-0.027}$ & $1.018^{+0.009}_{-0.010}$ & $0.882^{+0.044}_{-0.040}$ & $10.3^{+2.1}_{-2.7}$ & $0.191^{+0.067}_{-0.062}$ \\
WASP-43   & $4723^{+91}_{-82}$ & $4.579^{+0.019}_{-0.012}$ & $0.11^{+0.10}_{-0.10}$ & $0.763^{+0.024}_{-0.015}$ & $0.741^{+0.005}_{-0.006}$ & $0.246^{+0.018}_{-0.016}$ & $9.0^{+3.1}_{-4.7}$ & $1.06^{+0.12}_{-0.13}$ \\
WASP-173A & $5792^{+97}_{-69}$ & $4.381^{+0.042}_{-0.033}$ & $0.09^{+0.10}_{-0.11}$ & $1.026^{+0.070}_{-0.054}$ & $1.081^{+0.014}_{-0.016}$ & $1.184^{+0.066}_{-0.046}$ & $5.4^{+2.6}_{-3.1}$ & $0.074^{+0.070}_{-0.051}$ \\
\hline                                   
\end{tabular}
%\tablefoot{None?.}
\end{table*}
%------------------ END OF TABLE

\begin{figure*}
        \includegraphics[width=2\columnwidth]{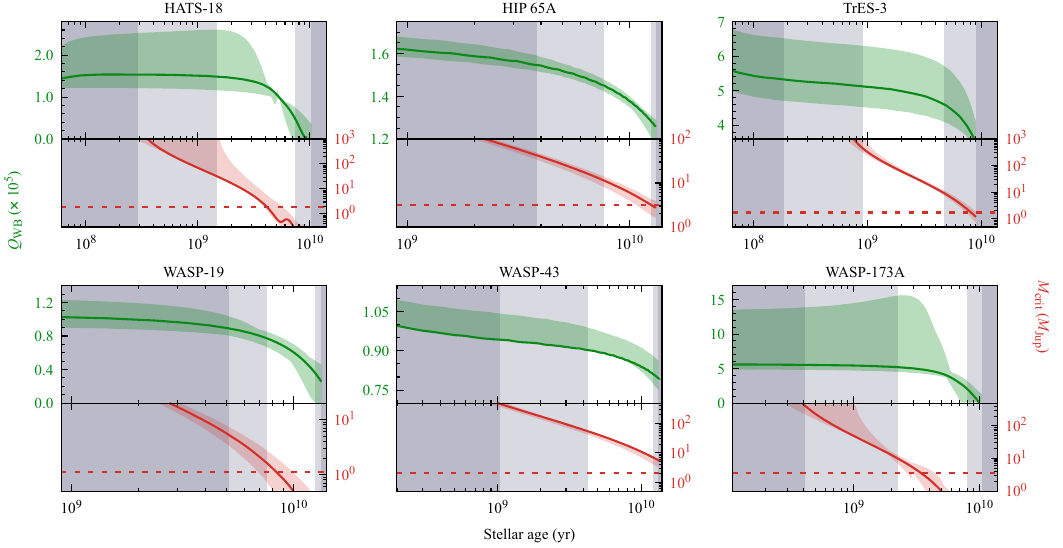}
    \caption{Theoretical behaviour of $Q'_{ \rm \star,WB}$ (green lines in upper panels) and $M_{\rm crit}$ (red lines in lower panels) as a function of stellar age for the planet-star systems of our sample. Green- and red-shaded areas reflect the propagation of 1 $\sigma$ uncertainties in the stellar parameters used as the inputs. The horizontal dashed lines mark the actual masses of the planets. Vertical grey shading codes the system ages with the 1, 2, and more $\sigma$ range marked in white, light grey, and dark grey, respectively. The values of $Q'_{ \rm \star,WB}$ are marked at the left-hand ordinates and $M_{\rm crit}$ at the right ones.}
    \label{fig:qmcrit}
\end{figure*}

To determine the expected tidal dissipation rate due to breaking the internal gravity waves, $Q'_{ \rm \star,WB}$, we started with constructing the evolutionary models of the host stars with MESA. The redetermined stellar masses and metallicities (Table~\ref{tab.StellarPars}) were used as the inputs. The MESA settings are given in Appendix~\ref{App:1}. We derived the stellar interior profiles for density, pressure, mass, and Brunt-V{\"a}is{\"a}l{\"a} frequency. We used these quantities to calculate the Eulerian gravitational potential perturbation, defined in Eq.~(8) by \citet{2020MNRAS.498.2270B}, by solving the second-order ordinary differential Eq.~(12) of \citet{2020MNRAS.498.2270B} with satisfying boundary conditions dictated by Eqs.~(13) and (14), which are also provided in that paper. In our calculations, we considered only the most important tide component for a coplanar configuration, characterised by the mode degree $l=2$ and the azimuthal order $m=2$. The potential perturbation was then used to calculate the radial component of the displacement field $\xi_{\rm r}$ following Eq.~(15) of \citet{2020MNRAS.498.2270B}. Using Brunt-V{\"a}is{\"a}l{\"a} frequency, we found the interface between the radiative and convective zones and calculated the local derivative of $\xi_{\rm r}$ and the logarithmic derivative of Brunt-V{\"a}is{\"a}l{\"a} frequency squared. Finally, we used Eq.~(41) of \citet{2020MNRAS.498.2270B}, fed with these quantities, the systemic parameters, and the parameters derived in Eqs.~(42) and (43) of \citet{2020MNRAS.498.2270B}, to calculate the theoretical value of $Q'_{ \rm \star,WB}$ for each evolutionary point in MESA tracks. The procedure was repeated for the stellar masses and metallicities increased and decreased by their 1~$\sigma$ uncertainties to trace the influence of these uncertainties on  $Q'_{ \rm \star,WB}$.

The results for the individual systems of our sample are shown in Fig.~\ref{fig:qmcrit}. The behaviour of $Q'_{ \rm \star,WB}$ as a function of stellar age is illustrated with green curves in the upper panels. The propagation of the 1~$\sigma$ uncertainties of the stellar parameters is marked with green-shaded areas. For the convenience of further discussion (Sect.~\ref{Sect:Discussion}), our redetermined stellar ages are coded with a grey scale with a 1 $\sigma$ range shown in white. The predicted values of $Q'_{ \rm \star,WB}$ within the 1 $\sigma$ range of the stellar parameters (the mass, metallicity, and age) are given for the individual systems in Table~\ref{tab.ques}.

Wave breaking is expected if waves are strong enough to overturn a radiative core's stratification. Following the criterion formulated in Eq. (47) by \citet{2020MNRAS.498.2270B}, we calculated a critical planetary mass $M_{\rm crit}$ required for wave breaking in the individual systems. The behaviour of this quantity as a function of the stellar age is plotted in lower panels in Fig.~\ref{fig:qmcrit} with red continuous lines. Red-shaded areas show the propagation of the 1~$\sigma$ uncertainties of the stellar parameters. To facilitate the discussion, we also marked the masses of the hot Jupiters with horizontal dashed red lines.

The magnitude of the tidal dissipation in the regime of weak nonlinearity, $Q'_{ \rm \star,WNL}$, was estimated by inter- and extrapolating the numerical results collected in Table~4 of \citet{2024ApJ...960...50W}. We employed the multivariate function {\tt interpn} implemented in the Python library {\tt SciPy} \citep{2020SciPy-NMeth}. The results for the individual systems are given in Table~\ref{tab.ques}.

\section{Discussion}\label{Sect:Discussion}

For most systems of our sample, the redetermined stellar parameters agree with the values reported in the literature. This study discusses only $M_{\star}$ and the age because these parameters were used in dynamical tides modelling. Our determinations of the masses for HATS-18, HIP 65A, TrES-3, WASP-19, and WASP-173A are consistent within the 1 $\sigma$ uncertainties with the previously reported values. Only for WASP-43, the discrepancies are substantial, reaching 3.9 $\sigma$. They are discussed in Sect.~\ref{Sect:Disc:W43}. Our stellar ages are in line with the literature determinations.

For HATS-18, TrES-3, WASP-19, and WASP-43, there are estimates for $Q'_{ \rm \star,WB}$ provided by \citet{2020MNRAS.498.2270B}. As summarised in Table~\ref{tab.ques}, they agree with our theoretical results. \citet{2024ApJ...960...50W} listed all our sample systems as favourable targets for the search for orbital decay due to the WNL effects. Below, we comment on the results for individual systems and, if needed, discuss discrepancies with the literature in detail.

\subsection{HATS-18}\label{Sect:Disc:H18}

The system comprises a solar-mass G-type dwarf and a 2 $M_{\rm Jup}$ planet on a 0.84 d orbit \citep{2016AJ....152..127P}. In the discovery paper, \citet{2016AJ....152..127P} identified the system as one of the best candidates for probing tidal planet-star interactions. \citet{2020MNRAS.498.2270B} named it a promising target for detecting orbital decay driven by the IGW dissipation. Wave breaking could be expected if the star is older than 4.6 Gyr, resulting in efficient tidal dissipation with $Q'_{ \rm \star,WB} \approx 1.1 \times 10^{5}$. \citet{2022MNRAS.515.3212S} performed a thorough transit timing analysis for the first time using new ground-based observations and TESS data from sectors 10 and 36. The time coverage of 5.4 years allowed those authors to place a lower constraint on $Q'_{\star}$ of $1.29^{+0.12}_{-0.11} \times 10^{5}$, comparable with the theoretical prediction, which was refined to $Q'_{ \rm \star,WB} \approx 1.2 \times 10^{5}$. With the new TESS observations in sector 63, we extend the timing data set by 2 years and push the lower limit on $Q'_{\star}$ by a factor of 3, that is, to $3.5^{+0.5}_{-0.7} \times 10^{5}$.

As shown in Fig.~\ref{fig:qmcrit}, our predictions on $Q'_{ \rm \star,WB}$ in the 1 $\sigma$ range of the stellar parameters (the mass, metallicity, and age) are $\sim$$(1.2 - 2.6) \times 10^{5}$ for models younger than $\sim$2 Gyr. They align with the earlier predictions of \citet{2020MNRAS.498.2270B} and \citet{2022MNRAS.515.3212S} up to the age of $\sim$$4$ Gy. The values of $Q'_{ \rm \star,WB}$ drop below $\sim$$0.8 \times 10^{5}$ for older models, including stellar evolution stages after leaving the main sequence. The estimates for $M_{\rm crit}$ drop below the mass of HATS-18~b if the system is older than $\sim$$4-6$ Gy. In this case, $Q'_{ \rm \star,WB}$ would be below $\sim$$1.3 \times 10^{5}$, and the orbital decay would be unequivocally detectable in the current timing data set. Wave breaking is not expected if the system is younger; our observations support this option.

\begin{figure}
        \includegraphics[width=\columnwidth]{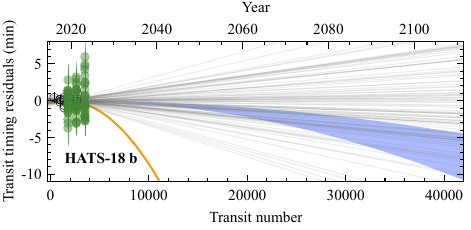}
    \caption{Predicted departure from the current linear transit ephemeris for HATS-18~b. The mustard line traces the dissipation through the weakly nonlinear effects for $Q'_{ \rm \star,WNL} \approx 5.5 \times 10^5$ derived from \citet{2024ApJ...960...50W}. A blue area illustrates predictions of the gyrochronological analysis by \citet{2016AJ....152..127P} with $Q'_{\star}$ between $1.0 \times 10^{7}$ (lower envelope) and $2.4 \times 10^{7}$ (upper envelope). The data points are the same as in Fig.~\ref{fig:tt}. The uncertainties of the linear ephemeris are illustrated with 100 lines drawn from the posterior distribution.}
    \label{fig:h18pred}
\end{figure}

With our limitation on $Q'_{\star}$, we enter the weakly nonlinear tidal dissipation regime. It is still below the predicted $Q'_{ \rm \star,WNL} \approx 5.5 \times 10^5$. We expect that precise transit timing acquired shortly (i.e. in a couple of years) will allow us to explore this regime fully. We used the predicted value of $Q'_{ \rm \star,WNL}$ to trace a cumulative transit time shift that could build up with time. As shown with the mustard line in Fig.~\ref{fig:h18pred}, a departure by 10 minutes against the present-day linear ephemeris might be reached in 24 years from $E=0$, that is, around the calendar year 2039. Naturally, a firm detection could happen much earlier, depending on the timing precision of follow-up observations\footnote{We recall the case of the WASP-12 system, in which the orbital decay of the ultra-hot Jupiter was detected when the departure from the linear ephemeris reached about 5 minutes \citep{2016A&A...588L...6M} and was verified at the departure of about 10 minutes \citep[e.g.][]{2018AcA....68..371M}.}.

\citet{2016AJ....152..127P} noticed that HATS-18 exhibits the rotational variability induced by starspots. It rotates within $\sim$$10$ days, which is too fast compared to its age from stellar evolution models. This finding could be interpreted as transferring the orbital angular momentum towards the stellar spin due to the tidal interaction. \citet{2018AJ....155..165P} predict the value of $Q'_{\star}$ must range from $\sim$$(1.0-2.4) \times 10^{7}$ to account for the stellar spin excess. This value is two orders of magnitude above our empirical constraint. The range of the cumulative time shift in transits observed in the future is illustrated with the blue-shaded area in Fig.~\ref{fig:h18pred}. It builds up much more gently, and the departure of 10 minutes might be observed between 2110 and 2160.

\subsection{HIP 65A}\label{Sect:Disc:Hip65}

In this system, a massive planet of $3.2$ $M_{\rm Jup}$ orbits a 0.8 $M_{\odot}$ K4 dwarf within almost 1 day \citep{2020AA...639A..76N}. The host is chromospherically active and exhibits a photometric variation due to starspots. It constitutes a visual binary system with an M dwarf, separated by about 250 au ($3 \farcs 95$ on the sky). The transit timing data set we analysed comes only from the TESS observations and spans 5 years. Thus, our observational constraint on $Q'_{\star}$ of $7.6^{+0.8}_{-0.7} \times 10^{4}$ is still relatively weak. A similar conclusion was reached by \citet{2024PSJ.....5..163A}, who obtained $Q'_{\star} \gtrapprox 2 \times 10^{5}$ at 99.7\% confidence.

As displayed in Fig.~\ref{fig:qmcrit}, our model predicts $Q'_{ \rm \star,WB}$ within the 1 $\sigma$ of the stellar parameters ranging $(1.2-1.5) \times 10^{5}$. This is still above the current observational constraint. We find, however, that the system's configuration disfavours the breaking of IGWs. The value of $M_{\rm crit}$ required for full IGW dissipation is up to an order of magnitude greater than the mass of HIP~65A~b. Higher-mass models allow for wave breaking only for ages above $\sim$10.5 Gy. The weakly nonlinear effects could dissipate the tidal energy with $Q'_{ \rm \star,WNL}$ of $\sim$$1.3 \times 10^{6}$, translating into the cumulative transit time shift of 10 minutes in about 60 years from $E=0$, that is, around 2080.

\citet{2020AA...639A..76N} found that the star rotates within $\sim$$13$ days\footnote{More recently, \citet{2024AJ....167....1W} analysed out-of-transit photometry from TESS and found a photometric modulation with a much shorter period of $\sim$6 days.}, translating into its gyrochronological age of $0.32^{+0.10}_{-0.06}$ Gyr. Compared to the reported evolutionary age of $4.1^{+4.3}_{-2.8}$ Gyr, the difference was interpreted as evidence for the tidal spinning-up of the star. The value of $Q'_{\star} > 10^8$ was proposed to account for this effect. With this prediction, the transit time shift of 10 minutes would be built up in at least 500 years.

\subsection{TrES-3}\label{Sect:Disc:Tres3}

This system comprises a 0.9 $M_{\odot}$ G dwarf and a 1.8 $M_{\rm Jup}$ planet on a 1.3 d orbit \citep{2007ApJ...663L..37O}. A hint for orbital decay of TrES-3~b was reported by \citet{2020AJ....160...47M}, who performed a homogeneous analysis based on literature data enhanced by 12 new transit light curves. Those authors found $\dot{P}_{\rm orb}$ to be consistent with $Q'_{\star} \approx 1 \times 10^{5}$. However, the significance of this finding was disqualifying low, just at 1.3 $\sigma$. In their subsequent work, \citet{2022AJ....164..198M} enhanced the transit timing data set with TESS observations from sectors 25, 26, and 40. In addition, amateur light curves were used to extend the time coverage of the observations. Those authors noticed a lengthening instead of a shortening of the orbital period. Again, the statistical significance of this finding was disqualifying low, just at 1.6~$\sigma$. However, the positive value of ${{\rm d} P_{\rm{orb}}}/{{\rm d} E}$ resulted in a tight constraint on $Q'_{\star}$ of $2.8 \times 10^{6}$. Our tests on the original timing data showed that skipping the amateur data lowers this quantity to $\sim$$5 \times 10^{5}$. In contrast, \citet{2024PSJ.....5..163A} placed a much weaker constraint with $Q'_{\star} \gtrapprox 3 \times 10^{4}$ at 99.7\% confidence.

Conversely, \citet{2023ApJS..265....4K} compiled all publicly available mid-transit times, including amateur observations, and obtained a non-zero negative $\dot{P}_{\rm orb}$ at the 4.9 $\sigma$ level. \citet{2023arXiv231017225W} analysed a set of mid-transit times from \citet{2022ApJS..259...62I} enhanced with amateur data to confirm the orbital decay with even higher statistical significance. However, those authors showed that a one-point-out test leaves $\dot{P}_{\rm orb}$ in a wide range of values, including reversal of the sign of the period change rate. In particular, they identified two precise transit mid-points from \citet{2017ApJ...848....9S} and \citet{2022NewA...9101680S}, whose original uncertainties could be significantly underestimated\footnote{The light curves from \citet{2022NewA...9101680S} were not qualified for our analysis due to their low quality.}.

We carefully investigated the literature light curves for this system to ensure our analysis is based only on reliable observations. \citet{2017ApJ...848....9S} acquired high-precision photometric time series on 12 March 2017 using a 3.5 m telescope equipped with a diffuser. While working with the original light curve declared in $\rm{BJD_{TDB}}$, we noticed that the redetermined mid-transit time occurred 70 s later than published in the original paper, making this point a 70 s outlier in a transit timing residual plot. This difference coincides with the UTC-to-TDB correction of 69.2 s. Thus, we concluded that the time system conversion must have been accidentally applied twice to the timestamps in the published light curve and made appropriate corrections to the original timestamps\footnote{Until the submission of this paper, we had not managed to confirm this finding.}. We also derived more realistic uncertainties, greater by a factor of 2.5 than the original value of 5 s.

A high-precision light curve analysed by \citet{2019AA...628A.115V} is another example of observations gathered through a diffuser. It was acquired with a 2 m telescope on 16 August 2018. The original mid-transit time and the mid-transit time redetermined by us using a final (originally detrended) light curve resulted in a 4 $\sigma$ positive outlying point in the plot of the transit timing residuals. As shown in Fig.~6 in the paper by \citet{2019AA...628A.115V}, the raw photometry lacks observations before the ingress. Also, some deformations in the ingress wing become exaggerated in the middle of the transit. The detrending procedure originally applied by \citet{2019AA...628A.115V} accounted for seeing, airmass, pointing drifts, total counts over flat and dark frames, and total counts in a sky ring around a photometric aperture. We reproduced airmass for each timestamp and carefully applied linear detrending against time and airmass to the raw data. We noticed that there were actually no data points before the beginning of the transit, and the observation run must have started just before the first contact. Those few before-transit measurements in the final light curve, plotted in Fig.~6 in \citet{2019AA...628A.115V}, were, in fact, part of the ingress wing, probably flattened by the originally applied detrending. Having just two parameters, we could not identify and remove the source of the deformation around the middle of the transit. Thus, we cut out those measurements. The ingress and egress wings were sufficient to derive a precise and reliable mid-transit time. After this processing, our redetermined mid-transit time was found to be on time predicted by the linear ephemeris within less than 1 $\sigma$. This exercise demonstrated that aggressive detrending of a light curve missing a solid out-of-transit flux base might yield precise but inaccurate results, and we recommend extreme caution while handling such data.

Our observational constraint on $Q'_{\star}$ of $2.50^{+0.12}_{-0.14} \times 10^{5}$ is still too low by a factor of $\sim$2 to probe the tidal dissipation due to wave breaking. As shown in Fig.~\ref{fig:qmcrit}, our simulations yield $Q'_{ \rm \star,WB}$ ranging $(4.2-6.4) \times 10^{5}$. These values fit $\sim$$6.5 \times 10^{5}$ derived by \citet{2020MNRAS.498.2270B}. This author also noted that wave breaking is, however, unlikely. Indeed, our calculations confirm this finding, showing that $M_{\rm crit}$ stays orders of magnitude greater than the planetary mass for the system's 1 $\sigma$ age range. The efficiency of the WNL dissipation is characterised by $Q'_{ \rm \star,WNL} \approx 2.5 \times 10^6$ and could produce a 10-minute departure from the linear ephemeris in 170 yr from $E=0$.

The host star exhibits a photometric modulation with a period of $\sim$9.3 days \citep{2024AJ....167....1W}. We used Eqs. (3) and (16) from \citet{2007ApJ...669.1167B} to calculate the gyrochronological age and its uncertainty\footnote{We used the $B-V$ colour index from \citet{2007ApJ...663L..37O}, de-reddened by $A_{\rm V}/3.1$.}. It was found to be equal to $0.55 \pm 0.08$ Gyr. Since this value is consistent with the evolutionary age of $2.6^{+2.1}_{-1.7}$ Gyr within 1.2 sigma, TrES-3 appears not to be noticeably spun up by its hot Jupiter. \citet{2018AJ....155..165P} predict $Q'_{\star} > 4 \times 10^6$ from this tidal spin-up, a value of the same order as $Q'_{ \rm \star,WNL}$.

\subsection{WASP-19}\label{Sect:Disc:W19}

In this system, a 0.9 $M_{\odot}$ G dwarf is orbited by a 1.1 $M_{\rm Jup}$ planet within 0.8 d \citep{2010ApJ...708..224H}. \citet{2020MNRAS.491.1243P} applied a uniform methodology to 74 ground-based transit observations spread over 10 years. Their transit timing data revealed no sign of a long-term trend that could be attributed to orbital decay. The authors placed the lower constraint on $Q'_{\star} > (1.23 \pm 0.23) \times 10^{6}$. A similar result was obtained by \citet{2022AA...668A.114R}, who included TESS observations from sectors 9 and 36. \citet{2024PSJ.....5..163A} incorporated transits from sectors 62 and 63 and also found no orbital decay, placing the lower constraint on $Q'_{\star} \gtrapprox 2.5 \times 10^{6}$ at 99.7\% confidence. \citet{2020AJ....160..155W} also did not detect a departure from the linear transit ephemeris. On the other hand, \citet{2020AJ....159..150P}, \citet{2023ApJS..265....4K}, and \citet{2023Univ...10...12K} reported on the detection of the orbital decay at the 5--7 $\sigma$ level.

Our analysis supports the non-detection of orbital decay and places the tightest constraint on $Q'_{\star}$ reported so far. \citet{2020MNRAS.498.2270B} predicts that wave breaking could occur if the host star were $\approx$$9$ Gyr old and could be absent for younger ages. Our simulations confirm this finding. As shown in Fig.~\ref{fig:qmcrit}, $M_{\rm crit}$ might be above the mass of WASP-19~b for ages up to about $10$ Gyr, coinciding with the nominal age of the star of $10.3^{+2.1}_{-2.7}$ Gyr. For older ages, wave breaking could be expected. The predicted values of $Q'_{ \rm \star,WB}$ in the 1 $\sigma$ range of the stellar parameters are $(0.7-0.9) \times 10^{5}$ for the young ages, which overlap with $(0.6-0.8) \times 10^{5}$ reported by \citet{2020MNRAS.498.2270B}. For aged models, $Q'_{ \rm \star,WB}$ might be below $\sim$$0.5 \times 10^{5}$. These predictions are more than an order of magnitude lower than our empirical limitations of $Q'_{\star} > 4.79^{+0.31}_{-0.29} \times 10^{6}$. We also probed the regime of weakly nonlinear tidal dissipation up to an order of magnitude above the theoretical prediction of $Q'_{ \rm \star,WNL}$ of about $4.2 \times 10^5$. We, therefore, notice that gravity waves are not efficiently damped either via wave breaking or weak nonlinear effects in the interior of WASP-19.

The rotational variability of the host star allowed \citet{2010ApJ...708..224H} to determine its rotation period, which was found to be about 10.5 days. Those authors estimated the gyrochronological age to be similar to that of the Hyades, that is, $\sim$0.6 Gyr. Our calculations based on \citet{2007ApJ...669.1167B} yield $0.97 \pm 0.15$ Gyr. It is much lower than the evolutionary age of $10.3^{+2.1}_{-2.7}$ Gyr obtained by us. This result suggests that WASP-19 might be noticeably spun up by its giant planet. \citet{2018AJ....155..165P} found spin-up $Q'_{\star}\approx (6.5-8.1) \times 10^{6}$. Our constraint is close to this range, and follow-up observations in the next decades might enable exploring this regime. 

Some ground-based transit light curves acquired by \citet{2013MNRAS.428.3671T} and \citet{2013MNRAS.436....2M} exhibit additional flux bumps attributed to the planet's crossing events over starspots along a transit chord. Our transit models did not account for starspot occultations, so we masked their photometric signatures in the literature light curves before redetermining the mid-transit times. The quality of TESS data was not high enough to allow us to identify such features in individual transits. We noted that the refined ephemeris has a reduced $\chi^2$ of $1.6$, suggesting that the uncertainties of the individual mid-transit times might be underestimated for both the linear and trial quadratic ephemerides. We attribute this noise excess to the stellar activity that could affect transit light curves, especially in ingress and egress \citep[e.g.][]{2013A&A...556A..19O}, and hence the mid-transit times.

\subsection{WASP-43}\label{Sect:Disc:W43}

This system comprises a 2.1 $M_{\rm Jup}$ planet on a 0.8 d orbit and a 0.8~$M_{\odot}$ K-type dwarf \citep{2011AA...535L...7H}. The star is chromospherically active, exhibiting strong \ion{Ca}{II} H and K emission and rotational photometric variability due to starspots \citep{2011ApJ...730L..31H}. The system was the subject of numerous transit timing studies. \citet{2014ApJ...781..116B}, \citet{2014A&A...563A..41M}, \citet{2016AJ....151...17J}, and \citet{2018ChAA..42..101S} reported on preliminary detection of rapid orbital decay. This finding was not confirmed in further works by \citet{2016AJ....151..137H}, \citet{2020AJ....159..150P}, \citet{2020AJ....160..155W}, \citet{2021AJ....162..210D}, and \citet{2024PSJ.....5..163A}. However, \citet{2021AJ....162..210D} reported a hint of a potential downward trend in transit timing of WASP-43~b, based on TESS data from sectors 9 and 35. We do not confirm this trend after adding transit mid-points from sector 62, which was observed 2 years afterwards. All TESS data do follow the linear ephemeris. We suppose that the finding of \citet{2021AJ....162..210D} was rendered by applying an inaccurate transit ephemeris, affected by the early epoch mid-transit times with the underestimated uncertainties.

Our constraint on $Q'_{\star}$ of $(1.13 \pm 0.05) \times 10^6$ is the tightest reported for this system so far. It is an order of magnitude greater than the theoretical predictions of the wave-breaking scenario. As illustrated in Fig.~\ref{fig:qmcrit}, our simulations yield $Q'_{ \rm \star,WB} \approx (0.8-1.0) \times 10^{5}$, slightly below  $\sim$$1.3 \times 10^{5}$ predicted by \citet{2020MNRAS.498.2270B}. The null detection of rapid orbital decay is not surprising because wave breaking is unlikely to operate in this system \citep{2020MNRAS.498.2270B}. We confirm this finding; $M_{\rm crit}$ stays much above the mass of WASP-43~b in the whole interval covered by our simulations, that is, 14 Gyr. Our limitation on $Q'_{\star}$ also exceeds the regime of the WNL dissipation by a factor of 2. We conclude that both dynamical tidal effects do not efficiently dissipate the tidal energy in WASP-43.

In the literature studies, the star was found to be less massive than our determination. \citet{2011AA...535L...7H} found two solutions with $0.58 \pm 0.05$ $M_{\odot}$ and $0.71 \pm 0.05$ $M_{\odot}$. The latter was confirmed by \citet{2012AA...542A...4G}, \citet{2014A&A...563A..40C}, and \citet{2022A&A...668A..17S}, who derived $0.717 \pm 0.025$ $M_{\odot}$, $0.713^{+0.018}_{-0.021}$ $M_{\odot}$, and $0.712^{+0.041}_{-0.036}$ $M_{\odot}$, respectively. \citet{2021AJ....162..210D} obtained $0.646^{+0.026}_{-0.025}$ $M_{\odot}$ through modelling the spectral energy distribution. Naturally, this spread in the stellar mass is related to divergent values of other parameters, such as the effective temperature, metallicity, and age. We leave a further discussion on that issue out of the scope of this study, referring to \citet{2021AJ....162..210D} for detailed deliberation. Our value of $0.763^{+0.024}_{-0.015}$ $M_{\odot}$ is, in turn, shifted towards the greater values. An evolutionary age between 2 and 13 Gyr, if the 1 $\sigma$ interval is considered, was reported in the literature. Our solution, $9.0^{+3.1}_{-4.7}$ Gyr, is shifted towards more advanced ages.

The rotation period of WASP-43 is 15.6 d, which gives a gyrochronological age of $400^{+200}_{-100}$ Myr \citep{2011AA...535L...7H}. This rotational youth suggests the planet significantly accelerated the stellar spin. However, the spin-up value of $Q'_{\star} \approx (2.6-6.1) \times 10^{7}$, calculated by \citet{2018AJ....155..165P}, is an order of magnitude above our observational constraint. This rate of orbital decay would produce a 10-minute shift of transits in at least 150 years.

\subsection{WASP-173A}\label{Sect:Disc:W173}

The system is constituted by a solar analogue (G3, $\sim$$1 M_{\odot}$) and a massive (3.7 $M_{\rm Jup}$) gas giant on a 1.4 d orbit \citep{2019MNRAS.482.1379H}. It is also known as KELT-22A, independently discovered by another group \citep{2019ApJS..240...13L}. The host star is the brighter component of a binary star system in the separation of $\sim$1400 au ($\sim$$6 \farcs 1$ on the sky). A photometric modulation produced by the stellar rotation and starspots was observed, but it remains unknown which component it can be attributed to \citep{2019MNRAS.482.1379H,2019ApJS..240...13L}.

\begin{figure}
        \includegraphics[width=\columnwidth]{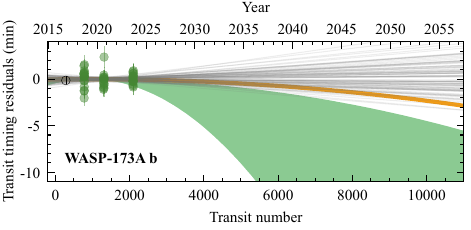}
    \caption{Predicted departure from the linear transit ephemeris for WASP-173A~b if dynamical tides are efficiently dissipated in the host's interior. The green area shows the possible scenarios for $Q'_{\rm \star,WB}$ between $1.5 \times 10^{5}$ (lower envelope, the present constraint from observations) and $1.5 \times 10^{6}$ (upper envelope) if wave breaking operates. The data points are the same as in Fig.~\ref{fig:tt}. The mustard line tracks the time shift governed by dissipation due to the weakly nonlinear effects with $Q'_{\rm \star,WNL}$ of $\sim 2.9 \times 10^{6}$ that is close to the value obtained from stellar spinning-up by \citet{2019ApJS..240...13L}. The uncertainties of the present linear ephemeris are illustrated with 100 lines drawn from the posterior distribution.}
    \label{fig:w173pred}
\end{figure}

As illustrated in Fig.~\ref{fig:qmcrit}, the predicted $Q'_{\rm \star,WB}$ ranges from $\sim$$5 \times 10^{5}$ to $\sim$$1.5 \times 10^{6}$ if the host star is 2--3 Gy old, and drops below $\sim$$4 \times 10^{5}$ at the end of the 1 $\sigma$ age range. However, if the host star is younger than $\sim$$4.5$ Gy, $M_{\rm crit}$ is above the mass of the hot Jupiter's mass, preventing from wave breaking. With our observational constraint of $Q'_{\star} > 1.50^{+0.19}_{-0.17} \times 10^{5}$, we enter the regime of those dynamical tides, eliminating a small fraction of the models with the lowest values of $Q'_{\rm \star,WB}$. As shown in Fig.~\ref{fig:w173pred}, the time shift for transits is expected to reach 10 minutes in 20 years from $E=0$, that is, around 2037 in the most favourite scenario. Thus, observations acquired in the upcoming decade can help verify these predictions. 

The rate of WNL dissipation is expected to be weaker with $Q'_{ \rm \star,WNL}$ of $\sim$$2.9 \times 10^{6}$. The observational effects are also predicted to be weaker. A mustard line in Fig.~\ref{fig:w173pred} shows the expected shortening of the $P_{\rm orb}$ for WASP-173A~b under this regime. A transit time shift of 10 minutes could be observed around 2090.

\citet{2019ApJS..240...13L} noted that the relatively short rotation period of WASP-173A, falling in a range of 7--9 days, might be caused by tidal spinning-up\footnote{The periodicity of the photometric variability is consistent with a stellar rotation period of $\sim$8 days derived from the projected radial velocity. Thus, it is reasonable to assume that the modulation is due to component A \citep{2019ApJS..240...13L}.}. In this scenario, the value of $Q'_{\star} \approx 10^{6.2} \approx 1.6 \times 10^6$ is predicted. Interestingly, this value coincides with $Q'_{\rm \star,WNL}$, rendering a unique opportunity for studying both effects in the future.

\section{Conclusions}\label{Sect:Conclusions}

We demonstrated that efficient dissipation of IGW energy due to wave breaking does not operate in the HATS-18 and WASP-19 planet-star configurations. In their evolution, these systems must still be at a stage before the wave-breaking regime, and the critical mass required for IGW breaking is higher than the planetary mass. Otherwise, orbital decay would be easily detected in the present transit timing data sets. This finding places upper constraints on the age of both host stars. HATS-18 must be younger than 4--6 Gyr and WASP-19 younger than 10 Gyr. Naturally, these deliberations are only valid if the expected magnitude of this IGW dissipation is not overestimated.

The WASP-173A system follows the analogous pattern. However, the time coverage of the timing measurements needs to be longer to probe this effect. Wave breaking is expected if the host star is older than $\sim$$4.5$ Gy. For younger stellar ages, the hot Jupiter is not massive enough to induce IGW breaking. The dissipation might be strong enough to produce observational effects in the following years or decades. Thus, we recognise WASP-173A~b as a promising candidate for further transit timing observations.

Our models show that wave breaking and strong dissipation under the IGW regime are unlikely to occur in the HIP 65A, TrES-3, and WASP-43 systems. That is because the ultra-hot Jupiters in those systems are not massive enough to trigger wave breaking at the current stellar ages. Our observational constraints for the first two systems still need to be tighter to verify this finding. The lower constraint on the dissipation rate in the WASP-43 system is one order of magnitude greater than that if IGWs were to break. Since this non-detection of orbital decay agrees with the theoretical expectations, it seems reasonable to assume that IGW breaking does not occur in HIP 65A and TrES-3 either.

The dissipation of dynamical tides under the WNL regime was probed in the WASP-19 and WASP-43 systems. The non-detection of orbital decay in those configurations shows that the theoretical dissipation rates might be overestimated by at least an order of magnitude. HATS-18, with its giant planet, is the most promising target among the remaining systems; our calculations show that precise transit timing could explore the WNL predictions in several years. For HIP 65A, WASP-173A, and TrES-3, much more extended coverage of decades or centuries is required to reach that goal. However, the null detection of WNL dissipation in the WASP-19 and WASP-43 systems is meaningful, suggesting that the theory might need some fine-tuning. It makes us doubt the possibility of this phenomenon occurring with the predicted magnitude in at least some of the remaining systems.

Gyrochronology may provide the actual estimates for the efficiency of tidal dissipation. The expected values of $Q'_{\star}$ in the HATS-18, HIP 65A, WASP-19, and WASP-43 systems are 1--2 orders of magnitude greater than those predicted by the WNL models. For TrES-3 and WASP-173A, those values coincide, suggesting that weakly non-linear dissipation, as modelled presently, could drive tidal interactions in those planet-star configurations. Whilst a century-long series of timing observations are required for TrES-3 to verify this finding, the goal can be achieved in several decades for the WASP-173A system.

\section{Data availability}\label{Sect:Data_availability}

The ground-based light curves and Table~\ref{tab.TTimes} in its entity are only available in electronic form at the CDS via anonymous ftp to cdsarc.u-strasbg.fr (130.79.128.5) or via http://cdsweb.u-strasbg.fr/cgi-bin/qcat?J/A+A/.

\begin{acknowledgements}
We thank the anonymous referee for careful reading and insightful comments that strengthened this manuscript.
We are in debt to Adrian Barker for discussing tidal dissipation calculations.
We thank Coel Hellier, Diana Dragomir, Romina Petrucci, and {\c{C}}a\u{g}lar P{\"u}sk{\"u}ll{\"u} for sharing the follow-up observations used in their studies.
We also thank all those authors who make their photometric data publicly available.
MF, VC, and DPM acknowledge financial support from the Agencia Estatal de Investigaci\'on (AEI/10.13039/501100011033) of the Ministerio de Ciencia e Innovaci\'on and the ERDF ''A way of making Europe'' through projects PID2022-137241NB-C43 and the Centre of Excellence ''Severo Ochoa'' award to the Instituto de Astrof\'{\i}sica de Andaluc\'{\i}a (CEX2021-001131-S). DPM also acknowledges grant AST\_00001\_8 of the financial entities UE(NextGenerationEU)-MICIU-PRTR-CSIC-JA.
This research is based on observations made at the Sierra Nevada Observatory (OSN), operated by the Institute of Astrophysics of Andalusia (IAA-CSIC).
Computations were carried out using the Hydra Cluster operating in the Institute of Astronomy, Nicolaus Copernicus University in Toru\'n.
This paper includes data collected with the TESS mission, obtained from the MAST data archive at the Space Telescope Science Institute (STScI). Funding for the TESS mission is provided by the NASA Explorer Program. STScI is operated by the Association of Universities for Research in Astronomy, Inc., under NASA contract NAS 5-26555. 
This research made use of Lightkurve, a Python package for Kepler and TESS data analysis \citep{2018ascl.soft12013L}. This research has made use of the SIMBAD database and the VizieR catalogue access tool, operated at CDS, Strasbourg, France, and NASA's Astrophysics Data System Bibliographic Services.
\end{acknowledgements}

\bibliographystyle{aa} % style aa.bst 
\bibliography{pap_iv} % your references Yourfile.bib

\begin{thebibliography}{96}
\expandafter\ifx\csname natexlab\endcsname\relax\def\natexlab#1{#1}\fi

\bibitem[{{Adams} {et~al.}(2024){Adams}, {Jackson}, {Sickafoose},
  {Morgenthaler}, {Worters}, {Stubbers}, {Carlson}, {Bhure}, {Dekeyser},
  {Huang}, \& {Weinberg}}]{2024PSJ.....5..163A}
{Adams}, E.~R., {Jackson}, B., {Sickafoose}, A.~A., {et~al.} 2024, PSJ, 5, 163

\bibitem[{{Barker}(2020)}]{2020MNRAS.498.2270B}
{Barker}, A.~J. 2020, \mnras, 498, 2270

\bibitem[{{Barker} \& {Ogilvie}(2010)}]{2010MNRAS.404.1849B}
{Barker}, A.~J. \& {Ogilvie}, G.~I. 2010, \mnras, 404, 1849

\bibitem[{{Barnes}(2007)}]{2007ApJ...669.1167B}
{Barnes}, S.~A. 2007, \apj, 669, 1167

\bibitem[{{Blecic} {et~al.}(2014){Blecic}, {Harrington}, {Madhusudhan},
  {Stevenson}, {Hardy}, {Cubillos}, {Hardin}, {Bowman}, {Nymeyer}, {Anderson},
  {Hellier}, {Smith}, \& {Collier Cameron}}]{2014ApJ...781..116B}
{Blecic}, J., {Harrington}, J., {Madhusudhan}, N., {et~al.} 2014, \apj, 781,
  116

\bibitem[{{Bonomo} {et~al.}(2017){Bonomo}, {Desidera}, {Benatti}, {Borsa},
  {Crespi}, {Damasso}, {Lanza}, {Sozzetti}, {Lodato}, {Marzari}, {Boccato},
  {Claudi}, {Cosentino}, {Covino}, {Gratton}, {Maggio}, {Micela}, {Molinari},
  {Pagano}, {Piotto}, {Poretti}, {Smareglia}, {Affer}, {Biazzo}, {Bignamini},
  {Esposito}, {Giacobbe}, {H{\'e}brard}, {Malavolta}, {Maldonado}, {Mancini},
  {Martinez Fiorenzano}, {Masiero}, {Nascimbeni}, {Pedani}, {Rainer}, \&
  {Scandariato}}]{2017AA...602A.107B}
{Bonomo}, A.~S., {Desidera}, S., {Benatti}, S., {et~al.} 2017, \aap, 602, A107

\bibitem[{{Carter} \& {Winn}(2009)}]{2009ApJ...704...51C}
{Carter}, J.~A. \& {Winn}, J.~N. 2009, \apj, 704, 51

\bibitem[{{Chen} {et~al.}(2023){Chen}, {Xie}, {Zhou}, {Dong}, {Yang}, {Zhu},
  {Liu}, {Huang}, {Xiang}, {Wang}, {Zheng}, {Luo}, {Zhang}, \&
  {Zhu}}]{2023PNAS..12004179C}
{Chen}, D.-C., {Xie}, J.-W., {Zhou}, J.-L., {et~al.} 2023, Proceedings of the
  National Academy of Science, 120, e2304179120

\bibitem[{{Chen} {et~al.}(2014){Chen}, {van Boekel}, {Wang}, {Nikolov},
  {Fortney}, {Seemann}, {Wang}, {Mancini}, \& {Henning}}]{2014A&A...563A..40C}
{Chen}, G., {van Boekel}, R., {Wang}, H., {et~al.} 2014, \aap, 563, A40

\bibitem[{{Choi} {et~al.}(2016){Choi}, {Dotter}, {Conroy}, {Cantiello},
  {Paxton}, \& {Johnson}}]{2016ApJ...823..102C}
{Choi}, J., {Dotter}, A., {Conroy}, C., {et~al.} 2016, \apj, 823, 102

\bibitem[{{Claret} \& {Bloemen}(2011)}]{2011AA...529A..75C}
{Claret}, A. \& {Bloemen}, S. 2011, \aap, 529, A75

\bibitem[{{Collins} {et~al.}(2017){Collins}, {Kielkopf}, {Stassun}, \&
  {Hessman}}]{2017AJ....153...77C}
{Collins}, K.~A., {Kielkopf}, J.~F., {Stassun}, K.~G., \& {Hessman}, F.~V.
  2017, \aj, 153, 77

\bibitem[{{Col{\'o}n} {et~al.}(2010){Col{\'o}n}, {Ford}, {Lee}, {Mahadevan}, \&
  {Blake}}]{2010MNRAS.408.1494C}
{Col{\'o}n}, K.~D., {Ford}, E.~B., {Lee}, B., {Mahadevan}, S., \& {Blake},
  C.~H. 2010, \mnras, 408, 1494

\bibitem[{{Cort{\'e}s-Zuleta} {et~al.}(2020){Cort{\'e}s-Zuleta}, {Rojo},
  {Wang}, {Hinse}, {Hoyer}, {Sanhueza}, {Correa-Amaro}, \&
  {Albornoz}}]{2020AA...636A..98C}
{Cort{\'e}s-Zuleta}, P., {Rojo}, P., {Wang}, S., {et~al.} 2020, \aap, 636, A98

\bibitem[{{Cutri} {et~al.}(2003){Cutri}, {Skrutskie}, {van Dyk}, {Beichman},
  {Carpenter}, {Chester}, {Cambresy}, {Evans}, {Fowler}, {Gizis}, {Howard},
  {Huchra}, {Jarrett}, {Kopan}, {Kirkpatrick}, {Light}, {Marsh}, {McCallon},
  {Schneider}, {Stiening}, {Sykes}, {Weinberg}, {Wheaton}, {Wheelock}, \&
  {Zacarias}}]{2003yCat.2246....0C}
{Cutri}, R.~M., {Skrutskie}, M.~F., {van Dyk}, S., {et~al.} 2003, {VizieR
  Online Data Catalog: 2MASS All-Sky Catalog of Point Sources (Cutri+ 2003)},
  VizieR On-line Data Catalog: II/246. Originally published in:
  2003yCat.2246....0C

\bibitem[{{Cutri} {et~al.}(2012){Cutri}, {Wright}, {Conrow}, \& {et
  al.}}]{2012yCat.2311....0C}
{Cutri}, R.~M., {Wright}, E.~L., {Conrow}, T., \& {et al.} 2012, {VizieR Online
  Data Catalog: WISE All-Sky Data Release (Cutri+ 2012)}, VizieR On-line Data
  Catalog: II/311. Originally published in: 2012wise.rept....1C

\bibitem[{{Davoudi} {et~al.}(2021){Davoudi}, {Ba{\c{s}}t{\"u}rk},
  {Yal{\c{c}}{\i}nkaya}, {Esmer}, \& {Safari}}]{2021AJ....162..210D}
{Davoudi}, F., {Ba{\c{s}}t{\"u}rk}, {\"O}., {Yal{\c{c}}{\i}nkaya}, S., {Esmer},
  E.~M., \& {Safari}, H. 2021, \aj, 162, 210

\bibitem[{{Dotter}(2016)}]{2016ApJS..222....8D}
{Dotter}, A. 2016, \apjs, 222, 8

\bibitem[{{Dragomir} {et~al.}(2011){Dragomir}, {Kane}, {Pilyavsky},
  {Mahadevan}, {Ciardi}, {Gazak}, {Gelino}, {Payne}, {Rabus}, {Ramirez}, {von
  Braun}, {Wright}, \& {Wyatt}}]{2011AJ....142..115D}
{Dragomir}, D., {Kane}, S.~R., {Pilyavsky}, G., {et~al.} 2011, \aj, 142, 115

\bibitem[{{Essick} \& {Weinberg}(2016)}]{2016ApJ...816...18E}
{Essick}, R. \& {Weinberg}, N.~N. 2016, \apj, 816, 18

\bibitem[{{Feroz} \& {Hobson}(2008)}]{2008MNRAS.384..449F}
{Feroz}, F. \& {Hobson}, M.~P. 2008, \mnras, 384, 449

\bibitem[{{Feroz} {et~al.}(2009){Feroz}, {Hobson}, \&
  {Bridges}}]{2009MNRAS.398.1601F}
{Feroz}, F., {Hobson}, M.~P., \& {Bridges}, M. 2009, \mnras, 398, 1601

\bibitem[{{Feroz} {et~al.}(2019){Feroz}, {Hobson}, {Cameron}, \&
  {Pettitt}}]{2019OJAp....2E..10F}
{Feroz}, F., {Hobson}, M.~P., {Cameron}, E., \& {Pettitt}, A.~N. 2019, The Open
  Journal of Astrophysics, 2, 10

\bibitem[{{Fulton} {et~al.}(2011){Fulton}, {Shporer}, {Winn}, {Holman},
  {P{\'a}l}, \& {Gazak}}]{2011AJ....142...84F}
{Fulton}, B.~J., {Shporer}, A., {Winn}, J.~N., {et~al.} 2011, \aj, 142, 84

\bibitem[{{Gaia Collaboration} {et~al.}(2023){Gaia Collaboration}, {Vallenari},
  {Brown}, {Prusti}, {de Bruijne}, {Arenou}, {Babusiaux}, {Biermann},
  {Creevey}, {Ducourant}, {Evans}, {Eyer}, {Guerra}, {Hutton}, {Jordi},
  {Klioner}, {Lammers}, {Lindegren}, {Luri}, {Mignard}, {Panem}, {Pourbaix},
  {Randich}, {Sartoretti}, {Soubiran}, {Tanga}, {Walton}, {Bailer-Jones},
  {Bastian}, {Drimmel}, {Jansen}, {Katz}, {Lattanzi}, {van Leeuwen}, {Bakker},
  {Cacciari}, {Casta{\~n}eda}, {De Angeli}, {Fabricius}, {Fouesneau},
  {Fr{\'e}mat}, {Galluccio}, {Guerrier}, {Heiter}, {Masana}, {Messineo},
  {Mowlavi}, {Nicolas}, {Nienartowicz}, {Pailler}, {Panuzzo}, {Riclet}, {Roux},
  {Seabroke}, {Sordo}, {Th{\'e}venin}, {Gracia-Abril}, {Portell}, {Teyssier},
  {Altmann}, {Andrae}, {Audard}, {Bellas-Velidis}, {Benson}, {Berthier},
  {Blomme}, {Burgess}, {Busonero}, {Busso}, {C{\'a}novas}, {Carry}, {Cellino},
  {Cheek}, {Clementini}, {Damerdji}, {Davidson}, {de Teodoro}, {Nu{\~n}ez
  Campos}, {Delchambre}, {Dell'Oro}, {Esquej}, {Fern{\'a}ndez-Hern{\'a}ndez},
  {Fraile}, {Garabato}, {Garc{\'\i}a-Lario}, {Gosset}, {Haigron}, {Halbwachs},
  {Hambly}, {Harrison}, {Hern{\'a}ndez}, {Hestroffer}, {Hodgkin}, {Holl},
  {Jan{\ss}en}, {Jevardat de Fombelle}, {Jordan}, {Krone-Martins}, {Lanzafame},
  {L{\"o}ffler}, {Marchal}, {Marrese}, {Moitinho}, {Muinonen}, {Osborne},
  {Pancino}, {Pauwels}, {Recio-Blanco}, {Reyl{\'e}}, {Riello}, {Rimoldini},
  {Roegiers}, {Rybizki}, {Sarro}, {Siopis}, {Smith}, {Sozzetti}, {Utrilla},
  {van Leeuwen}, {Abbas}, {{\'A}brah{\'a}m}, {Abreu Aramburu}, {Aerts},
  {Aguado}, {Ajaj}, {Aldea-Montero}, {Altavilla}, {{\'A}lvarez}, {Alves},
  {Anders}, {Anderson}, {Anglada Varela}, {Antoja}, {Baines}, {Baker},
  {Balaguer-N{\'u}{\~n}ez}, {Balbinot}, {Balog}, {Barache}, {Barbato},
  {Barros}, {Barstow}, {Bartolom{\'e}}, {Bassilana}, {Bauchet}, {Becciani},
  {Bellazzini}, {Berihuete}, {Bernet}, {Bertone}, {Bianchi}, {Binnenfeld},
  {Blanco-Cuaresma}, {Blazere}, {Boch}, {Bombrun}, {Bossini}, {Bouquillon},
  {Bragaglia}, {Bramante}, {Breedt}, {Bressan}, {Brouillet}, {Brugaletta},
  {Bucciarelli}, {Burlacu}, {Butkevich}, {Buzzi}, {Caffau}, {Cancelliere},
  {Cantat-Gaudin}, {Carballo}, {Carlucci}, {Carnerero}, {Carrasco},
  {Casamiquela}, {Castellani}, {Castro-Ginard}, {Chaoul}, {Charlot}, {Chemin},
  {Chiaramida}, {Chiavassa}, {Chornay}, {Comoretto}, {Contursi}, {Cooper},
  {Cornez}, {Cowell}, {Crifo}, {Cropper}, {Crosta}, {Crowley}, {Dafonte},
  {Dapergolas}, {David}, {David}, {de Laverny}, {De Luise}, {De March}, {De
  Ridder}, {de Souza}, {de Torres}, {del Peloso}, {del Pozo}, {Delbo},
  {Delgado}, {Delisle}, {Demouchy}, {Dharmawardena}, {Di Matteo}, {Diakite},
  {Diener}, {Distefano}, {Dolding}, {Edvardsson}, {Enke}, {Fabre}, {Fabrizio},
  {Faigler}, {Fedorets}, {Fernique}, {Fienga}, {Figueras}, {Fournier},
  {Fouron}, {Fragkoudi}, {Gai}, {Garcia-Gutierrez}, {Garcia-Reinaldos},
  {Garc{\'\i}a-Torres}, {Garofalo}, {Gavel}, {Gavras}, {Gerlach}, {Geyer},
  {Giacobbe}, {Gilmore}, {Girona}, {Giuffrida}, {Gomel}, {Gomez},
  {Gonz{\'a}lez-N{\'u}{\~n}ez}, {Gonz{\'a}lez-Santamar{\'\i}a},
  {Gonz{\'a}lez-Vidal}, {Granvik}, {Guillout}, {Guiraud},
  {Guti{\'e}rrez-S{\'a}nchez}, {Guy}, {Hatzidimitriou}, {Hauser}, {Haywood},
  {Helmer}, {Helmi}, {Sarmiento}, {Hidalgo}, {Hilger}, {H{\l}adczuk}, {Hobbs},
  {Holland}, {Huckle}, {Jardine}, {Jasniewicz}, {Jean-Antoine Piccolo},
  {Jim{\'e}nez-Arranz}, {Jorissen}, {Juaristi Campillo}, {Julbe}, {Karbevska},
  {Kervella}, {Khanna}, {Kontizas}, {Kordopatis}, {Korn}, {K{\'o}sp{\'a}l},
  {Kostrzewa-Rutkowska}, {Kruszy{\'n}ska}, {Kun}, {Laizeau}, {Lambert},
  {Lanza}, {Lasne}, {Le Campion}, {Lebreton}, {Lebzelter}, {Leccia}, {Leclerc},
  {Lecoeur-Taibi}, {Liao}, {Licata}, {Lindstr{\o}m}, {Lister}, {Livanou},
  {Lobel}, {Lorca}, {Loup}, {Madrero Pardo}, {Magdaleno Romeo}, {Managau},
  {Mann}, {Manteiga}, {Marchant}, {Marconi}, {Marcos}, {Marcos Santos},
  {Mar{\'\i}n Pina}, {Marinoni}, {Marocco}, {Marshall}, {Martin Polo},
  {Mart{\'\i}n-Fleitas}, {Marton}, {Mary}, {Masip}, {Massari},
  {Mastrobuono-Battisti}, {Mazeh}, {McMillan}, {Messina}, {Michalik}, {Millar},
  {Mints}, {Molina}, {Molinaro}, {Moln{\'a}r}, {Monari}, {Mongui{\'o}},
  {Montegriffo}, {Montero}, {Mor}, {Mora}, {Morbidelli}, {Morel}, {Morris},
  {Muraveva}, {Murphy}, {Musella}, {Nagy}, {Noval}, {Oca{\~n}a}, {Ogden},
  {Ordenovic}, {Osinde}, {Pagani}, {Pagano}, {Palaversa}, {Palicio},
  {Pallas-Quintela}, {Panahi}, {Payne-Wardenaar}, {Pe{\~n}alosa Esteller},
  {Penttil{\"a}}, {Pichon}, {Piersimoni}, {Pineau}, {Plachy}, {Plum}, {Poggio},
  {Pr{\v{s}}a}, {Pulone}, {Racero}, {Ragaini}, {Rainer}, {Raiteri}, {Rambaux},
  {Ramos}, {Ramos-Lerate}, {Re Fiorentin}, {Regibo}, {Richards}, {Rios Diaz},
  {Ripepi}, {Riva}, {Rix}, {Rixon}, {Robichon}, {Robin}, {Robin}, {Roelens},
  {Rogues}, {Rohrbasser}, {Romero-G{\'o}mez}, {Rowell}, {Royer}, {Ruz Mieres},
  {Rybicki}, {Sadowski}, {S{\'a}ez N{\'u}{\~n}ez}, {Sagrist{\`a} Sell{\'e}s},
  {Sahlmann}, {Salguero}, {Samaras}, {Sanchez Gimenez}, {Sanna},
  {Santove{\~n}a}, {Sarasso}, {Schultheis}, {Sciacca}, {Segol}, {Segovia},
  {S{\'e}gransan}, {Semeux}, {Shahaf}, {Siddiqui}, {Siebert}, {Siltala},
  {Silvelo}, {Slezak}, {Slezak}, {Smart}, {Snaith}, {Solano}, {Solitro},
  {Souami}, {Souchay}, {Spagna}, {Spina}, {Spoto}, {Steele},
  {Steidelm{\"u}ller}, {Stephenson}, {S{\"u}veges}, {Surdej}, {Szabados},
  {Szegedi-Elek}, {Taris}, {Taylor}, {Teixeira}, {Tolomei}, {Tonello}, {Torra},
  {Torra}, {Torralba Elipe}, {Trabucchi}, {Tsounis}, {Turon}, {Ulla}, {Unger},
  {Vaillant}, {van Dillen}, {van Reeven}, {Vanel}, {Vecchiato}, {Viala},
  {Vicente}, {Voutsinas}, {Weiler}, {Wevers}, {Wyrzykowski}, {Yoldas}, {Yvard},
  {Zhao}, {Zorec}, {Zucker}, \& {Zwitter}}]{2023A&A...674A...1G}
{Gaia Collaboration}, {Vallenari}, A., {Brown}, A.~G.~A., {et~al.} 2023, \aap,
  674, A1

\bibitem[{{Gazak} {et~al.}(2012){Gazak}, {Johnson}, {Tonry}, {Dragomir},
  {Eastman}, {Mann}, \& {Agol}}]{2012AdAst2012E..30G}
{Gazak}, J.~Z., {Johnson}, J.~A., {Tonry}, J., {et~al.} 2012, Advances in
  Astronomy, 2012, 697967

\bibitem[{{Gillon} {et~al.}(2012){Gillon}, {Triaud}, {Fortney}, {Demory},
  {Jehin}, {Lendl}, {Magain}, {Kabath}, {Queloz}, {Alonso}, {Anderson},
  {Collier Cameron}, {Fumel}, {Hebb}, {Hellier}, {Lanotte}, {Maxted},
  {Mowlavi}, \& {Smalley}}]{2012AA...542A...4G}
{Gillon}, M., {Triaud}, A.~H.~M.~J., {Fortney}, J.~J., {et~al.} 2012, \aap,
  542, A4

\bibitem[{{Hagey} {et~al.}(2022){Hagey}, {Edwards}, \&
  {Boley}}]{2022AJ....164..220H}
{Hagey}, S.~R., {Edwards}, B., \& {Boley}, A.~C. 2022, \aj, 164, 220

\bibitem[{{Hamer} \& {Schlaufman}(2019)}]{2019AJ....158..190H}
{Hamer}, J.~H. \& {Schlaufman}, K.~C. 2019, \aj, 158, 190

\bibitem[{{Hebb} {et~al.}(2010){Hebb}, {Collier-Cameron}, {Triaud}, {Lister},
  {Smalley}, {Maxted}, {Hellier}, {Anderson}, {Pollacco}, {Gillon}, {Queloz},
  {West}, {Bentley}, {Enoch}, {Haswell}, {Horne}, {Mayor}, {Pepe}, {Segransan},
  {Skillen}, {Udry}, \& {Wheatley}}]{2010ApJ...708..224H}
{Hebb}, L., {Collier-Cameron}, A., {Triaud}, A.~H.~M.~J., {et~al.} 2010, \apj,
  708, 224

\bibitem[{{Hellier} {et~al.}(2019){Hellier}, {Anderson}, {Bouchy}, {Burdanov},
  {Collier Cameron}, {Delrez}, {Gillon}, {Jehin}, {Lendl}, {Nielsen}, {Maxted},
  {Pepe}, {Pollacco}, {Queloz}, {S{\'e}gransan}, {Smalley}, {Triaud}, {Udry},
  \& {West}}]{2019MNRAS.482.1379H}
{Hellier}, C., {Anderson}, D.~R., {Bouchy}, F., {et~al.} 2019, \mnras, 482,
  1379

\bibitem[{{Hellier} {et~al.}(2011{\natexlab{a}}){Hellier}, {Anderson}, {Collier
  Cameron}, {Gillon}, {Jehin}, {Lendl}, {Maxted}, {Pepe}, {Pollacco}, {Queloz},
  {S{\'e}gransan}, {Smalley}, {Smith}, {Southworth}, {Triaud}, {Udry}, \&
  {West}}]{2011AA...535L...7H}
{Hellier}, C., {Anderson}, D.~R., {Collier Cameron}, A., {et~al.}
  2011{\natexlab{a}}, \aap, 535, L7

\bibitem[{{Hellier} {et~al.}(2011{\natexlab{b}}){Hellier}, {Anderson},
  {Collier-Cameron}, {Miller}, {Queloz}, {Smalley}, {Southworth}, \&
  {Triaud}}]{2011ApJ...730L..31H}
{Hellier}, C., {Anderson}, D.~R., {Collier-Cameron}, A., {et~al.}
  2011{\natexlab{b}}, \apjl, 730, L31

\bibitem[{{Hoyer} {et~al.}(2016){Hoyer}, {Pall{\'e}}, {Dragomir}, \&
  {Murgas}}]{2016AJ....151..137H}
{Hoyer}, S., {Pall{\'e}}, E., {Dragomir}, D., \& {Murgas}, F. 2016, \aj, 151,
  137

\bibitem[{{Ivshina} \& {Winn}(2022)}]{2022ApJS..259...62I}
{Ivshina}, E.~S. \& {Winn}, J.~N. 2022, \apjs, 259, 62

\bibitem[{{Jermyn} {et~al.}(2023){Jermyn}, {Bauer}, {Schwab}, {Farmer}, {Ball},
  {Bellinger}, {Dotter}, {Joyce}, {Marchant}, {Mombarg}, {Wolf}, {Sunny Wong},
  {Cinquegrana}, {Farrell}, {Smolec}, {Thoul}, {Cantiello}, {Herwig}, {Toloza},
  {Bildsten}, {Townsend}, \& {Timmes}}]{2023ApJS..265...15J}
{Jermyn}, A.~S., {Bauer}, E.~B., {Schwab}, J., {et~al.} 2023, \apjs, 265, 15

\bibitem[{{Jiang} {et~al.}(2016){Jiang}, {Lai}, {Savushkin}, {Mkrtichian},
  {Antonyuk}, {Griv}, {Hsieh}, \& {Yeh}}]{2016AJ....151...17J}
{Jiang}, I.-G., {Lai}, C.-Y., {Savushkin}, A., {et~al.} 2016, \aj, 151, 17

\bibitem[{{Kokori} {et~al.}(2023){Kokori}, {Tsiaras}, {Edwards}, {Jones},
  {Pantelidou}, {Tinetti}, {Bewersdorff}, {Iliadou}, {Jongen}, {Lekkas},
  {Nastasi}, {Poultourtzidis}, {Sidiropoulos}, {Walter}, {W{\"u}nsche},
  {Abraham}, {Agnihotri}, {Albanesi}, {Arce-Mansego}, {Arnot}, {Audejean},
  {Aumasson}, {Bachschmidt}, {Baj}, {Barroy}, {Belinski}, {Bennett}, {Benni},
  {Bernacki}, {Betti}, {Biagini}, {Bosch}, {Brandebourg}, {Br{\'a}t},
  {Bretton}, {Brincat}, {Brouillard}, {Bruzas}, {Bruzzone}, {Buckland},
  {Cal{\'o}}, {Campos}, {Carre{\~n}o}, {Carrion Rodrigo}, {Casali},
  {Casalnuovo}, {Cataneo}, {Chang}, {Changeat}, {Chowdhury}, {Ciantini},
  {Cilluffo}, {Coliac}, {Conzo}, {Correa}, {Coulon}, {Crouzet}, {Crow},
  {Curtis}, {Daniel}, {Dauchet}, {Dawes}, {Deldem}, {Deligeorgopoulos},
  {Dransfield}, {Dymock}, {Eenm{\"a}e}, {Esseiva}, {Evans}, {Falco},
  {Farf{\'a}n}, {Fern{\'a}ndez-Laj{\'u}s}, {Ferratfiat}, {Ferreira},
  {Ferretti}, {Fio{\l}ka}, {Fowler}, {Futcher}, {Gabellini}, {Gainey},
  {Gaitan}, {Gajdo{\v{s}}}, {Garc{\'\i}a-S{\'a}nchez}, {Garlitz}, {Gillier},
  {Gison}, {Gonzales}, {Gorshanov}, {Grau Horta}, {Grivas}, {Guerra},
  {Guillot}, {Haswell}, {Haymes}, {Hentunen}, {Hills}, {Hose}, {Humbert},
  {Hurter}, {Hynek}, {Irzyk}, {Jacobsen}, {Jannetta}, {Johnson},
  {J{\'o}{\'z}wik-Wabik}, {Kaeouach}, {Kang}, {Kiiskinen}, {Kim}, {Kivila},
  {Koch}, {Kolb}, {Ku{\v{c}}{\'a}kov{\'a}}, {Lai}, {Laloum}, {Lasota}, {Lewis},
  {Liakos}, {Libotte}, {Lomoz}, {Lopresti}, {Majewski}, {Malcher}, {Mallonn},
  {Mannucci}, {Marchini}, {Mari}, {Marino}, {Marino}, {Mario}, {Marquette},
  {Mart{\'\i}nez-Bravo}, {Ma{\v{s}}ek}, {Matassa}, {Michel}, {Michelet},
  {Miller}, {Miny}, {Molina}, {Mollier}, {Monteleone}, {Montigiani},
  {Morales-Aimar}, {Mortari}, {Morvan}, {Mugnai}, {Murawski}, {Naponiello},
  {Naudin}, {Naves}, {N{\'e}el}, {Neito}, {Neveu}, {Noschese},
  {{\"O}{\u{g}}men}, {Ohshima}, {Orbanic}, {Pace}, {Pantacchini}, {Paschalis},
  {Pereira}, {Peretto}, {Perroud}, {Phillips}, {Pintr}, {Pioppa}, {Plazas},
  {Poelarends}, {Popowicz}, {Purcell}, {Quinn}, {Raetz}, {Rees}, {Regembal},
  {Rocchetto}, {Rocci}, {Rockenbauer}, {Roth}, {Rousselot}, {Rubia}, {Ruocco},
  {Russo}, {Salisbury}, {Salvaggio}, {Santos}, {Savage}, {Scaggiante},
  {Sedita}, {Shadick}, {Silva}, {Sioulas}, {{\v{S}}koln{\'\i}k}, {Smith},
  {Smolka}, {Solmaz}, {Stanbury}, {Stouraitis}, {Tan}, {Theusner}, {Thurston},
  {Tifner}, {Tomacelli}, {Tomatis}, {Trnka}, {Tyl{\v{s}}ar}, {Valeau},
  {Vignes}, {Villa}, {Vives Sureda}, {Vora}, {Vra{\v{s}}t'{\'a}k}, {Walliang},
  {Wenzel}, {Wright}, {Zambelli}, {Zhang}, \&
  {Z{\'\i}bar}}]{2023ApJS..265....4K}
{Kokori}, A., {Tsiaras}, A., {Edwards}, B., {et~al.} 2023, \apjs, 265, 4

\bibitem[{{Korth} \& {Parviainen}(2023)}]{2023Univ...10...12K}
{Korth}, J. \& {Parviainen}, H. 2023, Universe, 10, 12

\bibitem[{{Kundurthy} {et~al.}(2013){Kundurthy}, {Becker}, {Agol}, {Barnes}, \&
  {Williams}}]{2013ApJ...764....8K}
{Kundurthy}, P., {Becker}, A.~C., {Agol}, E., {Barnes}, R., \& {Williams}, B.
  2013, \apj, 764, 8

\bibitem[{{Labadie-Bartz} {et~al.}(2019){Labadie-Bartz}, {Rodriguez},
  {Stassun}, {Ciardi}, {Penev}, {Johnson}, {Gaudi}, {Col{\'o}n}, {Bieryla},
  {Latham}, {Pepper}, {Collins}, {Evans}, {Relles}, {Siverd}, {Bento}, {Yao},
  {Stockdale}, {Tan}, {Zhou}, {Eastman}, {Albrow}, {Bayliss}, {Beatty},
  {Berlind}, {Bozza}, {Calkins}, {Cohen}, {Curtis}, {Esquerdo}, {Feliz},
  {Fulton}, {Gregorio}, {James}, {Jensen}, {Johnson}, {Johnson}, {Joner},
  {Kasper}, {Kielkopf}, {Kuhn}, {Lund}, {Malpas}, {Manner}, {McCrady},
  {McLeod}, {Oberst}, {Penny}, {Reed}, {Sliski}, {Stephens}, {Stevens},
  {Villanueva}, {Wittenmyer}, {Wright}, \& {Zambelli}}]{2019ApJS..240...13L}
{Labadie-Bartz}, J., {Rodriguez}, J.~E., {Stassun}, K.~G., {et~al.} 2019,
  \apjs, 240, 13

\bibitem[{{Lendl} {et~al.}(2013){Lendl}, {Gillon}, {Queloz}, {Alonso}, {Fumel},
  {Jehin}, \& {Naef}}]{2013A&A...552A...2L}
{Lendl}, M., {Gillon}, M., {Queloz}, D., {et~al.} 2013, \aap, 552, A2

\bibitem[{{Lightkurve Collaboration} {et~al.}(2018){Lightkurve Collaboration},
  {Cardoso}, {Hedges}, {Gully-Santiago}, {Saunders}, {Cody}, {Barclay}, {Hall},
  {Sagear}, {Turtelboom}, {Zhang}, {Tzanidakis}, {Mighell}, {Coughlin}, {Bell},
  {Berta-Thompson}, {Williams}, {Dotson}, \& {Barentsen}}]{2018ascl.soft12013L}
{Lightkurve Collaboration}, {Cardoso}, J. V. d.~M., {Hedges}, C., {et~al.}
  2018, {Lightkurve: Kepler and TESS time series analysis in Python}

\bibitem[{{Maciejewski} {et~al.}(2016){Maciejewski}, {Dimitrov},
  {Fern{\'a}ndez}, {Sota}, {Nowak}, {Ohlert}, {Nikolov}, {Bukowiecki}, {Hinse},
  {Pall{\'e}}, {Tingley}, {Kjurkchieva}, {Lee}, \& {Lee}}]{2016A&A...588L...6M}
{Maciejewski}, G., {Dimitrov}, D., {Fern{\'a}ndez}, M., {et~al.} 2016, \aap,
  588, L6

\bibitem[{{Maciejewski} {et~al.}(2018){Maciejewski}, {Fern{\'a}ndez},
  {Aceituno}, {Mart{\'\i}n-Ruiz}, {Ohlert}, {Dimitrov}, {Szyszka}, {von Essen},
  {Mugrauer}, {Bischoff}, {Michel}, {Mallonn}, {Stangret}, \&
  {Mo{\'z}dzierski}}]{2018AcA....68..371M}
{Maciejewski}, G., {Fern{\'a}ndez}, M., {Aceituno}, F., {et~al.} 2018, \actaa,
  68, 371

\bibitem[{{Maciejewski} {et~al.}(2022){Maciejewski}, {Fern{\'a}ndez}, {Sota},
  {Amado}, {Dimitrov}, {Nikolov}, {Ohlert}, {Mugrauer}, {Bischoff}, {Heyne},
  {Hildebrandt}, {Stenglein}, {Ar{\'e}valo}, {Neira}, {Riesco}, {S{\'a}nchez
  Mart{\'\i}nez}, \& {Verdugo}}]{2022A&A...667A.127M}
{Maciejewski}, G., {Fern{\'a}ndez}, M., {Sota}, A., {et~al.} 2022, \aap, 667,
  A127

\bibitem[{{Maciejewski} {et~al.}(2013){Maciejewski}, {Puchalski}, {Saral},
  {Derman}, {Kitze}, {Bukowiecki}, {Seeliger}, \&
  {Neuhaeuser}}]{2013IBVS.6082....1M}
{Maciejewski}, G., {Puchalski}, D., {Saral}, G., {et~al.} 2013, Information
  Bulletin on Variable Stars, 6082, 1

\bibitem[{{Mackebrandt} {et~al.}(2017){Mackebrandt}, {Mallonn}, {Ohlert},
  {Granzer}, {Lalitha}, {Garc{\'\i}a Mu{\~n}oz}, {Gibson}, {Lee}, {Sozzetti},
  {Turner}, {Va{\v{n}}ko}, \& {Strassmeier}}]{2017A&A...608A..26M}
{Mackebrandt}, F., {Mallonn}, M., {Ohlert}, J.~M., {et~al.} 2017, \aap, 608,
  A26

\bibitem[{{Mancini} {et~al.}(2013){Mancini}, {Ciceri}, {Chen}, {Tregloan-Reed},
  {Fortney}, {Southworth}, {Tan}, {Burgdorf}, {Calchi Novati}, {Dominik},
  {Fang}, {Finet}, {Gerner}, {Hardis}, {Hinse}, {J{\o}rgensen}, {Liebig},
  {Nikolov}, {Ricci}, {Sch{\"a}fer}, {Sch{\"o}nebeck}, {Skottfelt}, {Wertz},
  {Alsubai}, {Bozza}, {Browne}, {Dodds}, {Gu}, {Harps{\o}e}, {Henning},
  {Hundertmark}, {Jessen-Hansen}, {Kains}, {Kerins}, {Kjeldsen}, {Lund},
  {Lundkvist}, {Madhusudhan}, {Mathiasen}, {Penny}, {Prof}, {Rahvar}, {Sahu},
  {Scarpetta}, {Snodgrass}, \& {Surdej}}]{2013MNRAS.436....2M}
{Mancini}, L., {Ciceri}, S., {Chen}, G., {et~al.} 2013, \mnras, 436, 2

\bibitem[{{Mannaday} {et~al.}(2020){Mannaday}, {Thakur}, {Jiang}, {Sahu},
  {Joshi}, {Pandey}, {Joshi}, {Yadav}, {Su}, {Sariya}, {Yeh}, {Griv},
  {Mkrtichian}, {Shlyapnikov}, {Moskvin}, {Ignatov}, {Va{\v{n}}ko}, \&
  {P{\"u}sk{\"u}ll{\"u}}}]{2020AJ....160...47M}
{Mannaday}, V.~K., {Thakur}, P., {Jiang}, I.-G., {et~al.} 2020, \aj, 160, 47

\bibitem[{{Mannaday} {et~al.}(2022){Mannaday}, {Thakur}, {Southworth}, {Jiang},
  {Sahu}, {Mancini}, {Va{\v{n}}ko}, {Kundra}, {Gajdo{\v{s}}}, {A-thano},
  {Sariya}, {Yeh}, {Griv}, {Mkrtichian}, \&
  {Shlyapnikov}}]{2022AJ....164..198M}
{Mannaday}, V.~K., {Thakur}, P., {Southworth}, J., {et~al.} 2022, \aj, 164, 198

\bibitem[{{Miyazaki} \& {Masuda}(2023)}]{2023AJ....166..209M}
{Miyazaki}, S. \& {Masuda}, K. 2023, \aj, 166, 209

\bibitem[{{Morton}(2015)}]{2015ascl.soft03010M}
{Morton}, T.~D. 2015, {isochrones: Stellar model grid package}, Astrophysics
  Source Code Library, record ascl:1503.010

\bibitem[{{M{\"u}ller} {et~al.}(2013){M{\"u}ller}, {Huber}, {Czesla}, {Wolter},
  \& {Schmitt}}]{2013AA...560A.112M}
{M{\"u}ller}, H.~M., {Huber}, K.~F., {Czesla}, S., {Wolter}, U., \& {Schmitt},
  J.~H.~M.~M. 2013, \aap, 560, A112

\bibitem[{{Murgas} {et~al.}(2014){Murgas}, {Pall{\'e}}, {Zapatero Osorio},
  {Nortmann}, {Hoyer}, \& {Cabrera-Lavers}}]{2014A&A...563A..41M}
{Murgas}, F., {Pall{\'e}}, E., {Zapatero Osorio}, M.~R., {et~al.} 2014, \aap,
  563, A41

\bibitem[{{Nielsen} {et~al.}(2020){Nielsen}, {Brahm}, {Bouchy}, {Espinoza},
  {Turner}, {Rappaport}, {Pearce}, {Ricker}, {Vanderspek}, {Latham}, {Seager},
  {Winn}, {Jenkins}, {Acton}, {Bakos}, {Barclay}, {Barkaoui}, {Bhatti},
  {Brice{\~n}o}, {Bryant}, {Burleigh}, {Ciardi}, {Collins}, {Collins}, {Cooke},
  {Csubry}, {dos Santos}, {Eigm{\"u}ller}, {Fausnaugh}, {Gan}, {Gillon},
  {Goad}, {Guerrero}, {Hagelberg}, {Hart}, {Henning}, {Huang}, {Jehin},
  {Jenkins}, {Jord{\'a}n}, {Kielkopf}, {Kossakowski}, {Lavie}, {Law}, {Lendl},
  {de Leon}, {Lovis}, {Mann}, {Marmier}, {McCormac}, {Mori}, {Moyano},
  {Narita}, {Osip}, {Otegi}, {Pepe}, {Pozuelos}, {Raynard}, {Relles}, {Sarkis},
  {S{\'e}gransan}, {Seidel}, {Shporer}, {Stalport}, {Stockdale}, {Suc},
  {Tamura}, {Tan}, {Tilbrook}, {Ting}, {Trifonov}, {Udry}, {Vanderburg},
  {Wheatley}, {Wingham}, {Zhan}, \& {Ziegler}}]{2020AA...639A..76N}
{Nielsen}, L.~D., {Brahm}, R., {Bouchy}, F., {et~al.} 2020, \aap, 639, A76

\bibitem[{{O'Donovan} {et~al.}(2007){O'Donovan}, {Charbonneau}, {Bakos},
  {Mandushev}, {Dunham}, {Brown}, {Latham}, {Torres}, {Sozzetti}, {Kov{\'a}cs},
  {Everett}, {Baliber}, {Hidas}, {Esquerdo}, {Rabus}, {Deeg}, {Belmonte},
  {Hillenbrand}, \& {Stefanik}}]{2007ApJ...663L..37O}
{O'Donovan}, F.~T., {Charbonneau}, D., {Bakos}, G.~{\'A}., {et~al.} 2007,
  \apjl, 663, L37

\bibitem[{{Oshagh} {et~al.}(2013){Oshagh}, {Santos}, {Boisse}, {Bou{\'e}},
  {Montalto}, {Dumusque}, \& {Haghighipour}}]{2013A&A...556A..19O}
{Oshagh}, M., {Santos}, N.~C., {Boisse}, I., {et~al.} 2013, \aap, 556, A19

\bibitem[{{Patel} \& {Espinoza}(2022)}]{2022AJ....163..228P}
{Patel}, J.~A. \& {Espinoza}, N. 2022, \aj, 163, 228

\bibitem[{{Patra} {et~al.}(2020){Patra}, {Winn}, {Holman}, {Gillon},
  {Burdanov}, {Jehin}, {Delrez}, {Pozuelos}, {Barkaoui}, {Benkhaldoun},
  {Narita}, {Fukui}, {Kusakabe}, {Kawauchi}, {Terada}, {Bouma}, {Weinberg}, \&
  {Broome}}]{2020AJ....159..150P}
{Patra}, K.~C., {Winn}, J.~N., {Holman}, M.~J., {et~al.} 2020, \aj, 159, 150

\bibitem[{{Paxton} {et~al.}(2011){Paxton}, {Bildsten}, {Dotter}, {Herwig},
  {Lesaffre}, \& {Timmes}}]{2011ApJS..192....3P}
{Paxton}, B., {Bildsten}, L., {Dotter}, A., {et~al.} 2011, \apjs, 192, 3

\bibitem[{{Paxton} {et~al.}(2013){Paxton}, {Cantiello}, {Arras}, {Bildsten},
  {Brown}, {Dotter}, {Mankovich}, {Montgomery}, {Stello}, {Timmes}, \&
  {Townsend}}]{2013ApJS..208....4P}
{Paxton}, B., {Cantiello}, M., {Arras}, P., {et~al.} 2013, \apjs, 208, 4

\bibitem[{{Paxton} {et~al.}(2015){Paxton}, {Marchant}, {Schwab}, {Bauer},
  {Bildsten}, {Cantiello}, {Dessart}, {Farmer}, {Hu}, {Langer}, {Townsend},
  {Townsley}, \& {Timmes}}]{2015ApJS..220...15P}
{Paxton}, B., {Marchant}, P., {Schwab}, J., {et~al.} 2015, \apjs, 220, 15

\bibitem[{{Paxton} {et~al.}(2018){Paxton}, {Schwab}, {Bauer}, {Bildsten},
  {Blinnikov}, {Duffell}, {Farmer}, {Goldberg}, {Marchant}, {Sorokina},
  {Thoul}, {Townsend}, \& {Timmes}}]{2018ApJS..234...34P}
{Paxton}, B., {Schwab}, J., {Bauer}, E.~B., {et~al.} 2018, \apjs, 234, 34

\bibitem[{{Paxton} {et~al.}(2019){Paxton}, {Smolec}, {Schwab}, {Gautschy},
  {Bildsten}, {Cantiello}, {Dotter}, {Farmer}, {Goldberg}, {Jermyn}, {Kanbur},
  {Marchant}, {Thoul}, {Townsend}, {Wolf}, {Zhang}, \&
  {Timmes}}]{2019ApJS..243...10P}
{Paxton}, B., {Smolec}, R., {Schwab}, J., {et~al.} 2019, \apjs, 243, 10

\bibitem[{{Penev} {et~al.}(2018){Penev}, {Bouma}, {Winn}, \&
  {Hartman}}]{2018AJ....155..165P}
{Penev}, K., {Bouma}, L.~G., {Winn}, J.~N., \& {Hartman}, J.~D. 2018, \aj, 155,
  165

\bibitem[{{Penev} {et~al.}(2016){Penev}, {Hartman}, {Bakos}, {Ciceri}, {Brahm},
  {Bayliss}, {Bento}, {Jord{\'a}n}, {Csubry}, {Bhatti}, {de Val-Borro},
  {Espinoza}, {Zhou}, {Mancini}, {Rabus}, {Suc}, {Henning}, {Schmidt}, {Noyes},
  {L{\'a}z{\'a}r}, {Papp}, \& {S{\'a}ri}}]{2016AJ....152..127P}
{Penev}, K., {Hartman}, J.~D., {Bakos}, G.~{\'A}., {et~al.} 2016, \aj, 152, 127

\bibitem[{{Penev} {et~al.}(2012){Penev}, {Jackson}, {Spada}, \&
  {Thom}}]{2012ApJ...751...96P}
{Penev}, K., {Jackson}, B., {Spada}, F., \& {Thom}, N. 2012, \apj, 751, 96

\bibitem[{{Petrucci} {et~al.}(2020){Petrucci}, {Jofr{\'e}}, {G{\'o}mez Maqueo
  Chew}, {Hinse}, {Ma{\v{s}}ek}, {Tan}, \& {G{\'o}mez}}]{2020MNRAS.491.1243P}
{Petrucci}, R., {Jofr{\'e}}, E., {G{\'o}mez Maqueo Chew}, Y., {et~al.} 2020,
  \mnras, 491, 1243

\bibitem[{{P{\"u}sk{\"u}ll{\"u}} {et~al.}(2017){P{\"u}sk{\"u}ll{\"u}},
  {Soydugan}, {Erdem}, \& {Budding}}]{2017NewA...55...39P}
{P{\"u}sk{\"u}ll{\"u}}, {\c{C}}., {Soydugan}, F., {Erdem}, A., \& {Budding}, E.
  2017, \na, 55, 39

\bibitem[{{Ricci} {et~al.}(2015){Ricci}, {Ram{\'o}n-Fox}, {Ayala-Loera},
  {Michel}, {Navarro-Meza}, {Fox-Machado}, {Reyes-Ruiz}, {Brown Sevilla}, \&
  {Curiel}}]{2015PASP..127..143R}
{Ricci}, D., {Ram{\'o}n-Fox}, F.~G., {Ayala-Loera}, C., {et~al.} 2015, \pasp,
  127, 143

\bibitem[{{Ricker} {et~al.}(2015){Ricker}, {Winn}, {Vanderspek}, {Latham},
  {Bakos}, {Bean}, {Berta-Thompson}, {Brown}, {Buchhave}, {Butler}, {Butler},
  {Chaplin}, {Charbonneau}, {Christensen-Dalsgaard}, {Clampin}, {Deming},
  {Doty}, {De Lee}, {Dressing}, {Dunham}, {Endl}, {Fressin}, {Ge}, {Henning},
  {Holman}, {Howard}, {Ida}, {Jenkins}, {Jernigan}, {Johnson}, {Kaltenegger},
  {Kawai}, {Kjeldsen}, {Laughlin}, {Levine}, {Lin}, {Lissauer}, {MacQueen},
  {Marcy}, {McCullough}, {Morton}, {Narita}, {Paegert}, {Palle}, {Pepe},
  {Pepper}, {Quirrenbach}, {Rinehart}, {Sasselov}, {Sato}, {Seager},
  {Sozzetti}, {Stassun}, {Sullivan}, {Szentgyorgyi}, {Torres}, {Udry}, \&
  {Villasenor}}]{2015JATIS...1a4003R}
{Ricker}, G.~R., {Winn}, J.~N., {Vanderspek}, R., {et~al.} 2015, Journal of
  Astronomical Telescopes, Instruments, and Systems, 1, 014003

\bibitem[{{Ros{\'a}rio} {et~al.}(2022){Ros{\'a}rio}, {Barros}, {Demangeon}, \&
  {Santos}}]{2022AA...668A.114R}
{Ros{\'a}rio}, N.~M., {Barros}, S.~C.~C., {Demangeon}, O.~D.~S., \& {Santos},
  N.~C. 2022, \aap, 668, A114

\bibitem[{{Saeed} {et~al.}(2022){Saeed}, {Goderya}, \&
  {Chishtie}}]{2022NewA...9101680S}
{Saeed}, M.~I., {Goderya}, S.~N., \& {Chishtie}, F.~A. 2022, \na, 91, 101680

\bibitem[{{Saha} {et~al.}(2021){Saha}, {Chakrabarty}, \&
  {Sengupta}}]{2021AJ....162...18S}
{Saha}, S., {Chakrabarty}, A., \& {Sengupta}, S. 2021, \aj, 162, 18

\bibitem[{{Savitzky} \& {Golay}(1964)}]{1964AnaCh..36.1627S}
{Savitzky}, A. \& {Golay}, M.~J.~E. 1964, Analytical Chemistry, 36, 1627

\bibitem[{{Scandariato} {et~al.}(2022){Scandariato}, {Singh}, {Kitzmann},
  {Lendl}, {Brandeker}, {Bruno}, {Bekkelien}, {Benz}, {Gutermann}, {Maxted},
  {Bonfanti}, {Charnoz}, {Fridlund}, {Heng}, {Hoyer}, {Pagano}, {Persson},
  {Salmon}, {Van Grootel}, {Wilson}, {Asquier}, {Bergomi}, {Gambicorti},
  {Hasiba}, {Alibert}, {Alonso}, {Anglada}, {B{\'a}rczy}, {Barrado y
  Navascues}, {Barros}, {Baumjohann}, {Beck}, {Beck}, {Billot}, {Bonfils},
  {Broeg}, {Cabrera}, {Collier Cameron}, {Csizmadia}, {Davies}, {Deleuil},
  {Deline}, {Delrez}, {Demangeon}, {Demory}, {Erikson}, {Fortier}, {Fossati},
  {Gandolfi}, {Gillon}, {G{\"u}del}, {Isaak}, {Kiss}, {Laskar}, {Lecavelier des
  Etangs}, {Lovis}, {Magrin}, {Nascimbeni}, {Olofsson}, {Ottensamer},
  {Pall{\'e}}, {Parviainen}, {Peter}, {Piotto}, {Pollacco}, {Queloz},
  {Ragazzoni}, {Rando}, {Rauer}, {Ribas}, {Santos}, {S{\'e}gransan}, {Serrano},
  {Simon}, {Smith}, {Sousa}, {Steller}, {Szab{\'o}}, {Thomas}, {Udry}, {Ulmer},
  \& {Walton}}]{2022A&A...668A..17S}
{Scandariato}, G., {Singh}, V., {Kitzmann}, D., {et~al.} 2022, \aap, 668, A17

\bibitem[{{Southworth} {et~al.}(2022){Southworth}, {Barker}, {Hinse}, {Jongen},
  {Dominik}, {J{\o}rgensen}, {Longa-Pe{\~n}a}, {Sajadian}, {Snodgrass},
  {Tregloan-Reed}, {Bach-M{\o}ller}, {Bonavita}, {Bozza}, {Burgdorf}, {Figuera
  Jaimes}, {Helling}, {Hitchcock}, {Hundertmark}, {Khalouei}, {Korhonen},
  {Mancini}, {Peixinho}, {Rahvar}, {Rabus}, {Skottfelt}, \&
  {Spyratos}}]{2022MNRAS.515.3212S}
{Southworth}, J., {Barker}, A.~J., {Hinse}, T.~C., {et~al.} 2022, \mnras, 515,
  3212

\bibitem[{{Southworth} {et~al.}(2009){Southworth}, {Hinse}, {J{\o}rgensen},
  {Dominik}, {Ricci}, {Burgdorf}, {Hornstrup}, {Wheatley}, {Anguita}, {Bozza},
  {Novati}, {Harps{\o}e}, {Kj{\ae}rgaard}, {Liebig}, {Mancini}, {Masi},
  {Mathiasen}, {Rahvar}, {Scarpetta}, {Snodgrass}, {Surdej}, {Th{\"o}ne}, \&
  {Zub}}]{2009MNRAS.396.1023S}
{Southworth}, J., {Hinse}, T.~C., {J{\o}rgensen}, U.~G., {et~al.} 2009, \mnras,
  396, 1023

\bibitem[{{Sozzetti} {et~al.}(2009){Sozzetti}, {Torres}, {Charbonneau}, {Winn},
  {Korzennik}, {Holman}, {Latham}, {Laird}, {Fernandez}, {O'Donovan},
  {Mandushev}, {Dunham}, {Everett}, {Esquerdo}, {Rabus}, {Belmonte}, {Deeg},
  {Brown}, {Hidas}, \& {Baliber}}]{2009ApJ...691.1145S}
{Sozzetti}, A., {Torres}, G., {Charbonneau}, D., {et~al.} 2009, \apj, 691, 1145

\bibitem[{{Stefansson} {et~al.}(2017){Stefansson}, {Mahadevan}, {Hebb},
  {Wisniewski}, {Huehnerhoff}, {Morris}, {Halverson}, {Zhao}, {Wright},
  {O'rourke}, {Knutson}, {Hawley}, {Kanodia}, {Li}, {Hagen}, {Liu}, {Beatty},
  {Bender}, {Robertson}, {Dembicky}, {Gray}, {Ketzeback}, {McMillan}, \&
  {Rudyk}}]{2017ApJ...848....9S}
{Stefansson}, G., {Mahadevan}, S., {Hebb}, L., {et~al.} 2017, \apj, 848, 9

\bibitem[{{Sun} {et~al.}(2018){Sun}, {Ji}, \& {Dong}}]{2018ChAA..42..101S}
{Sun}, Z., {Ji}, J.-h., \& {Dong}, Y. 2018, \caa, 42, 101

\bibitem[{{Tejada Arevalo} {et~al.}(2021){Tejada Arevalo}, {Winn}, \&
  {Anderson}}]{2021ApJ...919..138T}
{Tejada Arevalo}, R.~A., {Winn}, J.~N., \& {Anderson}, K.~R. 2021, \apj, 919,
  138

\bibitem[{{Tregloan-Reed} {et~al.}(2013){Tregloan-Reed}, {Southworth}, \&
  {Tappert}}]{2013MNRAS.428.3671T}
{Tregloan-Reed}, J., {Southworth}, J., \& {Tappert}, C. 2013, \mnras, 428, 3671

\bibitem[{{Turner} {et~al.}(2013){Turner}, {Smart}, {Hardegree-Ullman},
  {Carleton}, {Walker-LaFollette}, {Crawford}, {Smith}, {McGraw}, {Small},
  {Rocchetto}, {Cunningham}, {Towner}, {Zellem}, {Robertson}, {Guvenen},
  {Schwarz}, {Hardegree-Ullman}, {Collura}, {Henz}, {Lejoly}, {Richardson},
  {Weinand}, {Taylor}, {Daugherty}, {Wilson}, \&
  {Austin}}]{2013MNRAS.428..678T}
{Turner}, J.~D., {Smart}, B.~M., {Hardegree-Ullman}, K.~K., {et~al.} 2013,
  \mnras, 428, 678

\bibitem[{{Va{\v{n}}ko} {et~al.}(2013){Va{\v{n}}ko}, {Maciejewski},
  {Jakub{\'\i}k}, {Krej{\v{c}}ov{\'a}}, {Budaj}, {Pribulla}, {Ohlert}, {Raetz},
  {Parimucha}, \& {Bukowiecki}}]{2013MNRAS.432..944V}
{Va{\v{n}}ko}, M., {Maciejewski}, G., {Jakub{\'\i}k}, M., {et~al.} 2013,
  \mnras, 432, 944

\bibitem[{Virtanen {et~al.}(2020)Virtanen, Gommers, Oliphant, Haberland, Reddy,
  Cournapeau, Burovski, Peterson, Weckesser, Bright, {van der Walt}, Brett,
  Wilson, Millman, Mayorov, Nelson, Jones, Kern, Larson, Carey, Polat, Feng,
  Moore, {VanderPlas}, Laxalde, Perktold, Cimrman, Henriksen, Quintero, Harris,
  Archibald, Ribeiro, Pedregosa, {van Mulbregt}, \& {SciPy 1.0
  Contributors}}]{2020SciPy-NMeth}
Virtanen, P., Gommers, R., Oliphant, T.~E., {et~al.} 2020, Nature Methods, 17,
  261

\bibitem[{{von Essen} {et~al.}(2019){von Essen}, {Stefansson}, {Mallonn},
  {Pursimo}, {Djupvik}, {Mahadevan}, {Kjeldsen}, {Freudenthal}, \&
  {Dreizler}}]{2019AA...628A.115V}
{von Essen}, C., {Stefansson}, G., {Mallonn}, M., {et~al.} 2019, \aap, 628,
  A115

\bibitem[{{Wang} \& {Espinoza}(2024)}]{2024AJ....167....1W}
{Wang}, G. \& {Espinoza}, N. 2024, \aj, 167, 1

\bibitem[{{Wang} \& {Ma}(2024)}]{2024ESS.....560501W}
{Wang}, W. \& {Ma}, B. 2024, in AAS/Division for Extreme Solar Systems
  Abstracts, Vol.~56, AAS/Division for Extreme Solar Systems Abstracts, 605.01

\bibitem[{{Wang} {et~al.}(2024){Wang}, {Zhang}, {Chen}, {Wang}, {Yu}, \&
  {Ma}}]{2023arXiv231017225W}
{Wang}, W., {Zhang}, Z., {Chen}, Z., {et~al.} 2024, \apjs, 270, 14

\bibitem[{{Wang} {et~al.}(2021){Wang}, {Wang}, {Wang}, {Wu}, {Rice}, {Zhou},
  {Hinse}, {Liu}, {Ma}, {Peng}, {Zhang}, {Yu}, {Zhou}, \&
  {Laughlin}}]{2021ApJS..255...15W}
{Wang}, X.-Y., {Wang}, Y.-H., {Wang}, S., {et~al.} 2021, \apjs, 255, 15

\bibitem[{{Weinberg} {et~al.}(2024){Weinberg}, {Davachi}, {Essick}, {Yu},
  {Arras}, \& {Belland}}]{2024ApJ...960...50W}
{Weinberg}, N.~N., {Davachi}, N., {Essick}, R., {et~al.} 2024, \apj, 960, 50

\bibitem[{{Wong} {et~al.}(2020){Wong}, {Shporer}, {Daylan}, {Benneke},
  {Fetherolf}, {Kane}, {Ricker}, {Vanderspek}, {Latham}, {Winn}, {Jenkins},
  {Boyd}, {Glidden}, {Goeke}, {Sha}, {Ting}, \&
  {Yahalomi}}]{2020AJ....160..155W}
{Wong}, I., {Shporer}, A., {Daylan}, T., {et~al.} 2020, \aj, 160, 155

\bibitem[{{Zahn}(1966)}]{1966AnAp...29..489Z}
{Zahn}, J.~P. 1966, Annales d'Astrophysique, 29, 489

\bibitem[{{Zahn}(1975)}]{1975A&A....41..329Z}
{Zahn}, J.~P. 1975, \aap, 41, 329

\end{thebibliography}

\onecolumn
\begin{appendix}

\section{MESA settings}\label{App:1}
We used MESA version number r23.05.1. We added {\tt brunt\_N2} and {\tt brunt\_frequency} to the {\tt profile\_columns.list} file, which we needed for calculations in addition to {\tt logR}, {\tt logRho}, {\tt logP} and {\tt mass}. We also added {\tt phase\_of\_evolution} to the {\tt history\_columns.list} file, which we used for discussion. In our {\tt inlist\_project} file we changed {\tt initial\_mass}, {\tt initial\_z} and {\tt max\_age} for individual models. The general settings used are:
\begin{verbatim}
initial_mass = mass
initial_z = z
MLT_option = 'Henyey'
max_age = age
history_interval = 1
profile_interval = 1
max_num_profile_models = 1000
max_years_for_timestep = 1.0d8
mesh_delta_coeff = 0.3
when_to_stop_rtol = 1d-6
when_to_stop_atol = 1d-6
max_model_number = 1000
\end{verbatim}
 
\section{Additional tables}\label{App:2}

%------------------ TABLE OBS.TESS
\begin{table*}[h]
\caption{Details of the TESS observations used.} 
\label{tab.ObsTESS}      
\centering                  
\begin{tabular}{l c c c c c c c}      
\hline\hline                
System & Sector  & Camera & CCD & from -- to (UT) & $N_{\rm{obs}}$ & PNR (ppth) & $N_{\rm{tr}}$ \\
\hline
HATS-18   & 10 & 1 & 1 & 2019 Mar 26 -- 2019 Apr 22 & 16135 & 12.3 & 25 \\
          & 36 & 1 & 2 & 2021 Mar 07 -- 2021 Apr 01 & 16928 & 9.73 & 26 \\
          & 63 & 1 & 2 & 2023 Mar 10 -- 2023 Apr 06 & 17486 & 9.45 & 27 \\
                                 \multicolumn{7}{r}{total:} & 78 \\
HIP 65A   &  1 & 2 & 2 & 2018 Jul 25 -- 2018 Aug 22 & 18279 & 1.33 & 25 \\
          &  2 & 2 & 1 & 2018 Aug 23 -- 2018 Sep 20 & 18300 & 1.25 & 26 \\
          & 28 & 2 & 2 & 2020 Jul 31 -- 2020 Aug 25 & 17284 & 1.40 & 24 \\
          & 29 & 2 & 1 & 2020 Aug 26 -- 2020 Sep 21 & 17929 & 1.49 & 26 \\
          & 68 & 2 & 2 & 2023 Jul 29 -- 2023 Aug 25 & 17586 & 1.48 & 21 \\
          & 69 & 2 & 1 & 2023 Aug 25 -- 2023 Sep 20 & 16550 & 1.43 & 24 \\
                                \multicolumn{7}{r}{total:} & 146 \\
TrES-3    & 25 & 2 & 4 & 2020 May 14 -- 2020 Jun 08 & 17246 & 3.59 & 18 \\
          & 26 & 2 & 3 & 2020 Jun 09 -- 2020 Jul 04 & 16941 & 3.37 & 18 \\
          & 40 & 2 & 3 & 2021 Jun 25 -- 2021 Jul 23 & 19611 & 3.54 & 22 \\
          & 52 & 2 & 4 & 2022 May 19 -- 2022 Jun 12 & 16749 & 3.25 & 18 \\
          & 53 & 2 & 3 & 2022 Jun 13 -- 2022 Jul 08 & 17306 & 3.31 & 18 \\
                                 \multicolumn{7}{r}{total:} & 94 \\
WASP-19   &  9 & 2 & 2 & 2019 Mar 01 -- 2019 Mar 25 & 16878 & 4.16 & 29 \\
          & 36 & 2 & 1 & 2021 Mar 07 -- 2021 Apr 01 & 17343 & 3.94 & 29 \\
          & 62 & 2 & 2 & 2023 Feb 12 -- 2023 Mar 10 & 18025 & 4.02 & 28 \\
          & 63 & 2 & 1 & 2023 Mar 10 -- 2023 Apr 06 & 18358 & 3.62 & 33 \\
                                \multicolumn{7}{r}{total:} & 119 \\
WASP-43   &  9 & 1 & 1 & 2019 Mar 01 -- 2019 Mar 25 & 16878 & 2.57 & 28 \\
          & 35 & 1 & 2 & 2021 Feb 09 -- 2021 Mar 06 & 13992 & 2.45 & 24 \\
          & 62 & 1 & 3 & 2023 Feb 12 -- 2023 Mar 10 & 18024 & 2.52 & 31 \\
                                 \multicolumn{7}{r}{total:} & 83 \\
WASP-173A &  2 & 1 & 1 & 2018 Aug 23 -- 2018 Sep 20 & 18300 & 1.66 & 18 \\
          & 29 & 1 & 1 & 2020 Aug 26 -- 2020 Sep 21 & 16653 & 1.78 & 15 \\
          & 69 & 1 & 1 & 2023 Aug 25 -- 2023 Sep 20 & 16696 & 1.68 & 16 \\
                                 \multicolumn{7}{r}{total:} & 49 \\
\hline                                   
\end{tabular}
\tablefoot{$N_{\rm{obs}}$ is the number of useful data points. PNR is the photometric noise rate  \citep{2011AJ....142...84F} in parts per thousand (ppth) of the normalised flux per minute of observation. $N_{\rm{tr}}$ is the number of complete transit light curves used in this study.}
\end{table*}
%------------------ END OF TABLE

%------------------ TABLE OBS
\begin{table*}[h]
\caption{Details of the ground-based observing runs for TrES-3.} 
\label{tab.Obs}      
\centering                  
\begin{tabular}{l l c c c c c c c}      
\hline\hline                
Date UT (Epoch)  & Telescope & Filter  & UT start--end &  $X$                      & $N_{\rm{obs}}$ & $t_{\rm{exp}}$ (s) & $\Gamma$ & PNR (ppth)\\
\hline
2016 Mar 09 (2504) & TRE 1.2 & $V$   & 00:39--04:03 & $1.88 \rightarrow 1.11$                  & 160 & 60 & 0.88 & 1.81 \\
2016 Apr 29 (2543) & TRE 1.2 & $V$   & 23:24--02:41 & $1.27 \rightarrow 1.02$                  & 141 & 60 & 0.88 & 1.47 \\
2016 May 02 (2546) & TRE 1.2 & $V$   & 21:05--00:58 & $1.87 \rightarrow 1.07$                 & 154 & 60 & 0.88 & 2.42 \\
2016 May 19 (2559) & TRE 1.2 & $V$   & 20:47--00:14 & $1.56 \rightarrow 1.05$                  & 180 & 60 & 0.88 & 1.90 \\
2016 Jul 18 (2605) & TRE 1.2 & $I$   & 22:51--02:11 & $1.06 \rightarrow 1.56$                  & 167 & 60 & 0.88 & 1.28 \\
2016 Aug 25 (2634) & TRE 1.2 & $I$   & 20:03--23:19 & $1.04 \rightarrow 1.46$                  & 169 & 60 & 0.88 & 1.28 \\
2020 Mar 24 (3634) & TOR 0.6 & clear & 00:15--03:14 & $1.43 \rightarrow 1.07$                  & 268 & 37 & 1.50 & 1.46 \\
2020 Aug 06 (3738) & TRE 1.2 & clear & 20:36--00:12 & $1.03 \rightarrow 1.37$                  & 173 & 50 & 1.03 & 1.01 \\
2021 May 19 (3957) & OSN 0.9 & clear & 21:57--01:37 & $1.56 \rightarrow 1.01$                  & 188 & 50 & 1.07 & 1.06 \\
2021 Jun 02 (3967) & OSN 0.9 & clear & 23:21--02:49 & $1.10 \rightarrow 1.00 \rightarrow 1.05$ & 482 & 20 & 2.31 & 0.97 \\
2023 Apr 24 (4496) & OSN 0.9 & clear & 22:53--02:34 & $1.94 \rightarrow 1.04$                  & 584 & 20 & 2.31 & 0.89 \\
2023 May 11 (4509) & OSN 0.9 & clear & 22:50--01:47 & $1.48 \rightarrow 1.03$                  & 463 & 20 & 2.31 & 1.14 \\
2023 Jun 27 (4545) & OSN 0.9 & clear & 22:51--02:23 & $1.02 \rightarrow 1.00 \rightarrow 1.16$ & 551 & 20 & 2.31 & 0.94 \\
2024 Apr 25 (4777) & OSN 1.5 & clear & 00:02--03:32 & $1.41 \rightarrow 1.00$ & 236 & 40 & 1.32 & 1.39 \\
2024 May 15 (4793) & OSN 1.5 & clear & 21:37--00:52 & $1.80 \rightarrow 1.06$ & 330 & 30 & 1.85 & 0.84 \\
2024 May 15 (4793) & TOR 0.6 & clear & 21:04--00:50 & $1.37 \rightarrow 1.04$ & 322 & 40 & 1.46 & 1.41 \\
2024 May 29 (4803) & OSN 1.5 & clear & 22:39--02:23 & $1.22 \rightarrow 1.00 \rightarrow 1.01$ & 379 & 30 & 1.69 & 0.64 \\
2024 Jun 01 (4806) & OSN 1.5 & clear & 20:53--00:10 & $1.63 \rightarrow 1.03$                  & 311 & 30 & 1.69 & 0.87 \\
2024 Jun 11 (4813) & OSN 1.5 & clear & 00:24--03:45 & $1.00 \rightarrow 1.22$ & 338 & 30 & 1.69 & 0.78 \\
2024 Jul 01 (4829) & OSN 0.9 & clear & 22:07--01:05 & $1.03 \rightarrow 1.00 \rightarrow 1.06$ & 295 & 30 & 1.68 & 0.64 \\
2024 Aug 01 (4852) & OSN 0.9 & clear & 22:30--02:08  & $1.02 \rightarrow 1.73$ & 506 & 20 & 2.34 & 1.19 \\
\hline                                   
\end{tabular}
\tablefoot{Date UT is given for mid-points of the transits. Epoch is the transit number from the initial ephemeris given in the discovery papers. $X$~tracks the target's air mass during a run. $N_{\rm{obs}}$ is the number of useful scientific exposures. $t_{\rm{exp}}$ is the exposure time used. $\Gamma$ is the median number of exposures per minute. PNR is defined in Table~\ref{tab.ObsTESS}.}
\end{table*}
%------------------ END OF TABLE

%------------------ TABLE results
\begin{table*}[h]
\caption{Systemic parameters refined from the transit light curves observed with TESS.} 
\label{tab.results}      
\centering                  
\begin{tabular}{l c c c c c l}      
\hline\hline                
& $i_{\rm{b}}$ $(^{\circ})$ & $a_{\rm{b}}/R_{\star}$ & $R_{\rm{b}}/R_{\star}$ & $u_{\rm 1,TESS}$ & $u_{\rm 2,TESS}$ & Source \\
\hline
\multicolumn{7}{l}{HATS-18 b} \\
& $86.1^{+2.5}_{-2.3}$ & $3.71^{+0.10}_{-0.14}$ & $0.1289 \pm 0.0021$ & $0.47^{+0.16}_{-0.17}$ & $0.01^{+0.36}_{-0.31}$ & this paper \\
& $85.5^{+1.9}_{-2.8}$ & $3.71^{+0.11}_{-0.22}$ & $0.1347 \pm 0.0019$ & $...$ & $...$ & \citet{2016AJ....152..127P} \\
& $88.2 \pm 1.8$ & $3.807 \pm 0.052$ & $0.13426 \pm 0.00095$ & $...$ & $...$ & \citet{2022MNRAS.515.3212S} \\
& $...$ & $3.78^{+0.09}_{-0.14}$ & $0.1456^{+0.0021}_{-0.0024}$ & $0.20^{+0.16}_{-0.13}$ & $0.10^{+0.25}_{-0.19}$ & \citet{2022AJ....163..228P} \\
\multicolumn{7}{l}{HIP 65A b} \\
& $78.7^{+0.7}_{-1.0}$ & $5.30^{+0.11}_{-0.10}$ & $0.178^{+0.058}_{-0.031}$ & $0.50^{\rm {a}}$ & $0.23^{\rm {a}}$ & this paper \\
& $77.18^{+0.92}_{-1.00}$ & $5.289^{+0.051}_{-0.045}$ & $0.287^{+0.088}_{-0.068}$ & $0.545 \pm 0.037$ & $0.195 \pm 0.041$ & \citet{2020AA...639A..76N} \\
& $76.3^{+1.5}_{-1.3}$ & $5.180^{+0.089}_{-0.084}$ & $0.30^{+0.12}_{-0.10}$ & $0.504^{+0.053}_{-0.051}$ & $0.168^{+0.053}_{-0.046}$ & \citet{2020AJ....160..155W} \\
\multicolumn{7}{l}{TrES-3 b} \\
& $81.42^{+0.13}_{-0.16}$ & $5.724^{+0.053}_{-0.058}$ & $0.1575^{+0.0024}_{-0.0017}$ & $0.31^{\rm {a}}$ & $0.27^{\rm {a}}$ & this paper \\
& $81.96^{+0.13}_{-0.13}$ & $5.948^{+0.072}_{-0.073}$ & $0.16483^{+0.00076}_{-0.00081}$ & $...$ & $...$ & \citet{2021AJ....162...18S} \\
& $...$ & $5.82^{+0.12}_{-0.13}$ & $0.1706^{+0.0058}_{-0.0039}$ & $0.45^{+0.41}_{-0.30}$ & $0.11^{+0.35}_{-0.33}$ & \citet{2022AJ....163..228P} \\
\multicolumn{7}{l}{WASP-19 b} \\
& $79.49^{+0.47}_{-0.38}$ & $3.560^{+0.044}_{-0.041}$ & $0.1446 \pm 0.0018$ & $0.20^{+0.24}_{-0.14}$ & $0.44^{+0.26}_{-0.37}$ & this paper \\
& $78.76 \pm 0.13$ & $3.4521 \pm 0.0077$ & $0.14259 \pm 0.00023$ & $...$ & $...$ & \citet{2013MNRAS.436....2M} \\
& $79.08^{+0.34}_{-0.37}$ & $3.533^{+0.048}_{-0.052}$ & $0.14410^{+0.00049}_{-0.00050}$ & $...$ & $...$ & \citet{2020AA...636A..98C} \\
& $79.59^{+0.67}_{-0.59}$ & $3.582^{+0.074}_{-0.067}$ & $0.1522^{+0.0015}_{-0.0021}$ & $0.20^{+0.17}_{-0.14}$ & $0.38^{+0.24}_{-0.25}$ & \citet{2020AJ....160..155W} \\
& $79.17^{+0.33}_{-0.31}$ & $3.533^{+0.039}_{-0.037}$ & $0.14541^{+0.00070}_{-0.00077}$ & $0.403^{+0.061}_{-0.062}$ & $0.118^{+0.084}_{-0.082}$ & \citet{2022AA...668A.114R} \\
& $...$ & $3.60^{+0.06}_{-0.06}$ & $0.1519^{+0.0018}_{-0.0020}$ & $0.24^{+0.22}_{-0.16}$ & $0.38^{+0.26}_{-0.35}$ & \citet{2022AJ....163..228P} \\
\multicolumn{7}{l}{WASP-43 b} \\
& $81.99 \pm 0.16$ & $4.814^{+0.042}_{-0.041}$ & $0.1586^{+0.0020}_{-0.0017}$ & $0.36^{+0.26}_{-0.22}$ & $0.31^{+0.35}_{-0.38}$ & this paper \\
& $82.33 \pm 0.20$ & $4.918^{+0.053}_{-0.051}$ & $0.15945^{+0.00076}_{-0.00077}$ & $...$ & $...$ & \citet{2012AA...542A...4G} \\
& $82.11 \pm 0.10$ & $4.867 \pm 0.023$ & $0.15942 \pm 0.00041$ & $...$ & $...$ & \citet{2016AJ....151..137H} \\
& $81.65^{+0.29}_{-0.25}$ & $4.734^{+0.054}_{-0.053}$ & $0.1595^{+0.0015}_{-0.0020}$ & $0.32^{+0.16}_{-0.20}$ & $0.33^{+0.32}_{-0.24}$ & \citet{2020AJ....160..155W} \\
& $82.18 \pm 0.11$ & $4.854^{+0.026}_{-0.025}$ & $0.15805^{+0.00034}_{-0.00035}$ & $0.426 \pm 0.037$ & $0.275 \pm 0.040$ & \citet{2021AJ....162..210D} \\
& $...$ & $4.72^{+0.05}_{-0.05}$ & $0.1615^{+0.0017}_{-0.0025}$ & $0.55^{+0.22}_{-0.31}$ & $0.00^{+0.46}_{-0.30}$ & \citet{2022AJ....163..228P} \\
\multicolumn{7}{l}{WASP-173A b} \\
& $83.22^{+0.73}_{-0.59}$ & $4.452^{+0.110}_{-0.094}$ & $0.0901 \pm 0.0009$ & $0.39^{+0.15}_{-0.14}$ & $0.03^{+0.23}_{-0.23}$ & this paper \\
& $85.2 \pm 1.1$ & $4.78 \pm 0.17$ & $0.1109 \pm 0.0009$ & $...$ & $...$ & \citet{2019MNRAS.482.1379H} \\
& $86.5^{+2.2}_{-2.3}$ & $4.91^{+0.21}_{-0.33}$ & $0.1203^{+0.0032}_{-0.0028}$ & $...$ & $...$ & \citet{2019ApJS..240...13L} \\
& $89.2^{+2.3}_{-2.2}$ & $5.13^{+0.06}_{-0.12}$ & $0.11273^{+0.00085}_{-0.00088}$ & $0.236^{+0.067}_{-0.091}$ & $0.18^{+0.19}_{-0.13}$ & \citet{2020AJ....160..155W} \\
\hline                                   
\end{tabular}
\tablefoot{$^{\rm {a}}$ Interpolated from the theoretical tables of \citet{2011AA...529A..75C} and varied under the Gaussian penalty of the width of 0.1 for $u_{\rm 1,TESS}$ and 0.2 for $u_{\rm 2,TESS}$.}
\end{table*}
%------------------ END OF TABLE

%------------------ TABLE TTimes
\begin{table*}[h]
\caption{Mid-transit times reported in this paper.} 
\label{tab.TTimes}      
\centering                  
\begin{tabular}{l c c c l}      
\hline\hline                
Planet & $T_{\rm{mid}}$ $(\rm{BJD_{TDB}})$ & $+\sigma$ (d) & $-\sigma$ (d)  &  Data source\\
\hline
HATS-18 b & 2458572.049503 & 0.001749 & 0.001907 & TESS \\ 
HATS-18 b & 2458572.890064 & 0.001626 & 0.001626 & TESS \\ 
HATS-18 b & 2458573.727592 & 0.001291 & 0.001301 & TESS \\ 
\hline                                   
\end{tabular}
\tablefoot{This table is available in its entirety in a machine-readable form at the CDS. A portion is shown here for guidance regarding its form and content.}
\end{table*}
%------------------ END OF TABLE

%------------------ TABLE STARs
\begin{table*}[h]
\caption{Spectroscopic parameters for the host stars, taken from the literature as the inputs in this study.} 
\label{tab.Stars}      
\centering                  
\begin{tabular}{l c c c l c l}      
\hline\hline                
Star    & $T_{\rm eff}$ (K) & $\log g_{\star}$ (cgs) & [Fe/H] & Source & $K_{\rm{b}}$ $({\rm m\,s^{-1}})$ & Source\\
\hline
HATS-18 & $5600 \pm 120$ & $...$$^{\rm {a}}$ & $0.28\pm0.08$ & \citet{2016AJ....152..127P} & $415.2 \pm 10.0$ & \citet{2016AJ....152..127P} \\
HIP 65A & $4590 \pm 49$ & $...$$^{\rm {a}}$ & $0.18\pm0.08$ & \citet{2020AA...639A..76N} & $753.7 \pm 5.0$ & \citet{2020AA...639A..76N} \\
TrES-3 & $5650 \pm 75$ & $4.4 \pm 0.1$ & $-0.19\pm0.08$ & \citet{2009ApJ...691.1145S} & $348.5 \pm 4.9$ & \citet{2017AA...602A.107B} \\
WASP-19 & $5500 \pm 100$ & $4.5 \pm 0.2$ & $0.02 \pm 0.09$ & \citet{2010ApJ...708..224H} & $257.9 \pm 2.9$ & \citet{2017AA...602A.107B} \\
WASP-43 & $4400 \pm 200$ & $4.5 \pm 0.2$ & $-0.05\pm 0.17$ & \citet{2011AA...535L...7H} & $549.5 \pm 3.8$ & \citet{2021AJ....162..210D} \\
WASP-173A & $5700 \pm 150$ & $4.5 \pm 0.2$ & $0.16 \pm 0.14$ & \citet{2019MNRAS.482.1379H} & $645 \pm 7$ & \citet{2019MNRAS.482.1379H} \\
\hline                                   
\end{tabular}
\tablefoot{$T_{\rm eff}$ is the effective temperature. $\log g_{\star}$ is the gravity acceleration at the bottom of the photosphere. [Fe/H] is the metallicity. $K_{\rm{b}}$ is the amplitude of the reflex RV induced by the hot Jupiter companion. \\ $^{\rm {a}}$ A value from pure spectroscopic analysis is not provided in the source paper.}
\end{table*}
%------------------ END OF TABLE

\end{appendix}
\end{document}